\documentclass[aps,prb,a4paper,twocolumn,10pt,floatfix,showpacs,showkeys,amsmath,amssymb,nobibnotes,noeprint,unsortedaddress,superscriptaddress]{revtex4-2}

\setcitestyle{super}
\usepackage{graphicx}
\usepackage[export]{adjustbox}
\usepackage[dvipsnames]{xcolor}
\usepackage[utf8]{inputenc} 
\usepackage[T1]{fontenc} 
\usepackage{amsmath}
\usepackage{hyperref}
\usepackage[noabbrev,capitalise]{cleveref} 
\usepackage{csquotes}
\usepackage{xr}
\usepackage{upgreek}
\usepackage{caption}
\usepackage{subcaption}
\usepackage[percent]{overpic}
\usepackage{physics}
\usepackage{xspace}
\usepackage{comment}
\usepackage{blindtext}
\usepackage{verbatim}
\usepackage[explicit]{titlesec}

\graphicspath{{figures/}}

\renewcommand{\thesection}{\arabic{section}}

\titleformat{\section}{\large\bfseries\filcenter}{\thesection.}{1em}{#1}

\newcommand{\parencite}[1]{\citenum{#1}}
\newcommand{\jgen}{J_\mathrm{gen}}
\newcommand{\jrec}{J_\mathrm{rec}}

\newcommand{\jsc}{J_\mathrm{sc}}

\newcommand{\jmpp}{J_\mathrm{mpp}}
\newcommand{\Voc}{V_\mathrm{oc}}
\newcommand{\Vimp}{V_\mathrm{imp}}
\newcommand{\Vext}{V_\mathrm{ext}}
\newcommand{\Vmpp}{V_\mathrm{mpp}}
\newcommand{\Vtr}{\Delta V_\mathrm{tr}}
\newcommand{\Vbi}{V_\mathrm{bi}}
\newcommand{\FF}{F\!F}
\newcommand{\pFF}{pF\!F}
\newcommand{\FFmax}{F\!F_\mathrm{max}}
\newcommand{\Rs}{R_\mathrm{s}}
\newcommand{\Rp}{R_\mathrm{p}}
\newcommand{\Rtr}{R_\mathrm{tr}}
\newcommand{\Rext}{R_\mathrm{ext}}
\newcommand{\Rphoto}{R_\mathrm{photo}}
\newcommand{\Rtotal}{R_\mathrm{p,total}}
\newcommand{\nid}{n_\mathrm{id}}
\newcommand{\nsig}{n_\mathrm{\sigma}}

\newcommand{\nn}{n_\mathrm{n}}
\newcommand{\napp}{n_\mathrm{app}}

\newcommand{\EF}{\Delta E_\mathrm{F}}
\newcommand{\etacol}{\eta_{\mathrm{col}}}

\newcommand{\etadiss}{\eta_\mathrm{diss}}
\newcommand{\mueff}{\mu_\mathrm{eff}}
\newcommand{\sigmaeff}{\sigma_\mathrm{eff}}
\newcommand{\sigmaoc}{\sigma_\mathrm{oc}}
\newcommand{\sigmasc}{\sigma_\mathrm{sc}}
\newcommand{\bl}{\left(}

\newcommand{\br}{\right)}

\newcommand{\E}{\,\cdot\, 10^}
\newcommand{\kT}{k_B T}
\newcommand{\voc}{v_\mathrm{oc}}
\newcommand{\vimp}{\Vimp}

\newcommand{\der}{\mathrm{d}}

\begin{document}

\title{Transport resistance dominates the fill factor losses in record organic solar cells} 

\author{Chen Wang}
\affiliation{Institut für Physik, Technische Universität Chemnitz, 09126 Chemnitz, Germany}

\author{Roderick C. I. MacKenzie}
\affiliation{Department of Engineering, Durham University, Lower Mount Joy, South Road, Durham, DH1 3LE, UK}

\author{Uli Würfel}
\affiliation{Fraunhofer Institute for Solar Energy Systems ISE, Heidenhofstr. 2, 79110 Freiburg, Germany}
\affiliation{Freiburg Materials Research Center FMF, University of Freiburg, Stefan-Meier-Str. 21, 79104 Freiburg, Germany}

\author{Dieter Neher}
\affiliation{Institute of Physics and Astronomy, University of Potsdam, Karl-Liebknecht-Str.24-25, D-14476 Potsdam-Golm, Germany}

\author{Thomas Kirchartz}
\affiliation{IMD-3 Photovoltaik Forschungszentrum Jülich 52425 Jülich, Germany}
\affiliation{Faculty of Engineering and CENIDE University of Duisburg-Essen Carl-Benz-Str. 199, 47057 Duisburg, Germany}

\author{Carsten Deibel}
\email[email: ]{deibel@physik.tu-chemnitz.de}
\affiliation{Institut für Physik, Technische Universität Chemnitz, 09126 Chemnitz, Germany}

\author{Maria Saladina}
\email[email: ]{maria.saladina@physik.tu-chemnitz.de}
\affiliation{Institut für Physik, Technische Universität Chemnitz, 09126 Chemnitz, Germany}

\begin{abstract}
Organic photovoltaics are a promising solar cell technology well-suited to mass production using roll-to-roll processes. The efficiency of lab-scale solar cells has exceeded 20\% and considerable attention is currently being given to understanding and minimising the remaining loss mechanisms preventing higher efficiencies. While recent efficiency improvements are partly owed to reducing non-radiative recombination losses at open-circuit, the low fill factor due to a significant transport resistance is becoming \textit{the} Achilles heel of organic photovoltaics. The term transport resistance refers to a voltage and light intensity dependent charge collection loss in low-mobility materials. In this Perspective, we demonstrate that even the highest efficiency organic solar cells reported to-date have significant performance losses that can be attributed to transport resistance and that lead to high fill factor losses. We provide a closer look at the transport resistance and the material properties influencing it. We describe how to experimentally characterise and quantify the transport resistance by providing easy to follow instructions. Furthermore, the causes and theory behind transport resistance are detailed. In particular, we integrate the relevant figures of merit and different viewpoints on the transport resistance. Finally, we outline strategies that can be followed to minimise these charge collection losses in future solar cells.
\end{abstract}

\maketitle

\setcounter{secnumdepth}{3}
\setcounter{tocdepth}{3}

\section{Introduction}

Photovoltaics are becoming an increasingly important contribution to the global electricity supply. In combination with wind, they must cover the lion's share of electricity generation in future societies if carbon dioxide emission is to be minimised.\cite{haegel_photovoltaics_2023} While silicon solar cells are the economically dominant technology, emerging mass-production-compatible technologies are being actively investigated in both academia and industry. Among the emerging photovoltaic technologies, organic photovoltaics (OPV) requires the lowest material volume to almost fully absorb sunlight, and also offers the largest choice in material systems. Due to its low weight and the ability to tune the absorption spectrum to match the solar spectrum, OPV is currently considered for applications such as building-integrated PV and Agri-PV.\cite{ravishankar2020achieving,meitzner2021agrivoltaics,hu2022multifaceted}

Recently, organic solar cells have made significant progress in efficiencies, primarily due to the development of highly absorbing non-fullerene acceptor molecules. This development has been vital as it has led to progress in mitigating two major loss mechanisms, namely the high voltage losses at open circuit and insufficient photocurrent collection. The most significant remaining drawback of OPV material systems is that the charge carrier mobilities are significantly lower than in many inorganic solar cell materials. For instance, the difference to crystalline silicon is approximately 6 to 7 orders of magnitude. While the absorber layers can be very thin (100~nm or less) due to the high absorption coefficient of organic semiconductors, a consequence of the low mobilities is that the absorber layers \emph{have} to be as thin to collect all photogenerated charge carriers. Nevertheless, despite the reduced thickness, charge carrier collection is still imperfect and a major loss mechanism leading to a reduced fill factor ($\FF$). 

Imperfect charge carrier transport within the absorber layer essentially causes a resistance that is related to the density of charge carriers and that is therefore a function of both voltage and light intensity (i.e., the two factors that affect the charge carrier density). This resistance of the absorber layer is typically referred to as a transport resistance as it is caused by the finite conductivity of the absorber rather than the series resistance of other layers such as the electrodes. \cite{cheyns_analytical_2008,muller_analysis_2013,schiefer_determination_2014,wurfel_impact_2015,saladina_transport_2024} 

A closer look reveals that even in the most efficient cells (i.e., freshly prepared \enquote{hero} cells) -- where non-radiative voltage losses attracted most community attention and are regarded as the primary factor limiting efficiency compared to the radiative limit for single junction solar cells -- transport resistance losses are also very significant. As an example we consider the single junction organic solar cell with one of the highest certified efficiency values for \emph{binary} donor--acceptor blends of 19.1\%, published in March 2024.\cite{liu202419EfficiencyBinary} The fill factor of these record cells, with an average (in-house measured) power conversion efficiency of 19.3\%, is 79.6\%. From the data provided, we estimated\cite{saladina_transport_2024} the pseudo-fill factor -- the fill factor of the solar cell \emph{if it had neither internal (transport) nor external series resistance losses at all} -- to be 87.4\%. The difference of 7.8~percentage points is due to transport resistance losses even in these fresh record solar cells. 

In general, (i) even for fresh record organic solar cells the transport resistance \emph{is} an important loss mechanism, and (ii) when operating organic solar cells for a longer time, a performance drop dominated by a fill factor loss is often observed, which is due to an increase in transport resistance.\cite{woepke2022} It is therefore very important to understand the transport resistance as it significantly affects the performance and longevity of organic solar cells.

When considering how to quantify transport losses, researchers face a specific problem related to the breakdown of the superposition principle \cite{robinson_departures_1994}. The superposition principle refers to the idea that the current--voltage curve of a solar cell can be separated into a term for recombination that is equivalent to the dark current and one term with opposite sign for the photogenerated current. In solar cells where transport losses in the absorber layer are significant, the recombination current will become a strong function of how quickly the carriers are extracted. The faster the extraction happens, the lower the concentration of charge carriers will be at any given point on the $JV$ curve. Lower carrier concentration will result in lower recombination rates. Thus, one has to identify ways modifying the analytical equation of the $JV$ curve to include both a voltage and light intensity dependent modification to the recombination current. 

\begin{figure}[!tb]
    \centering
    \includegraphics[width=0.95\linewidth]{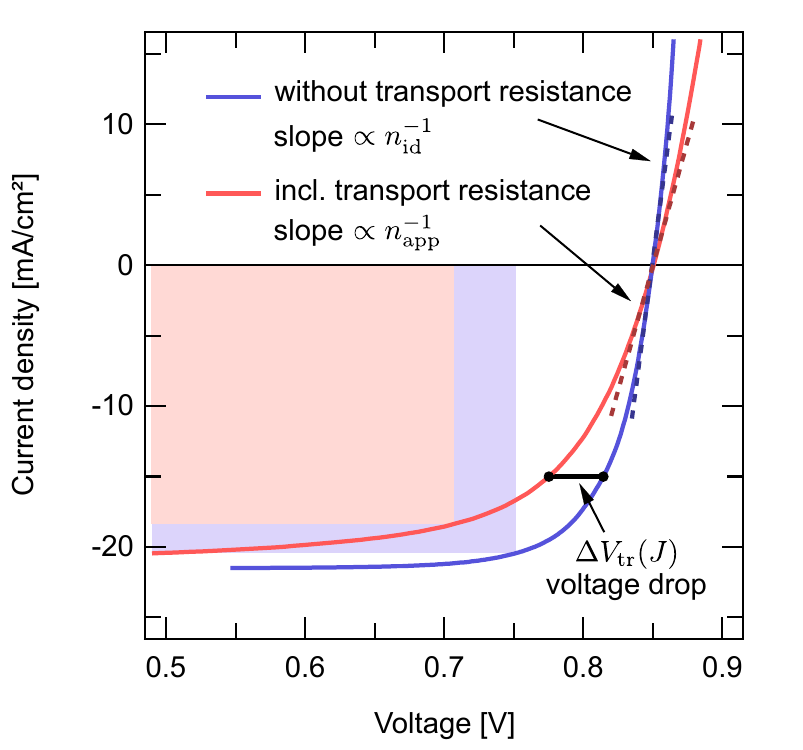}
    \caption{Current--voltage characteristics of a solar cell with and without transport resistance, with the shaded areas indicating the maximum output power. Transport resistance causes a voltage drop, $\Vtr$, which flattens the slope of the curve (corresponding to an increase in the apparent ideality factor) and leads to additional $\FF$ losses.}
    \label{fig:JV_intro}
\end{figure}

As mentioned earlier, one way to account for transport losses is to think of slow charge transport as an internal resistance within the material that makes it harder for charge carriers to move. Figure~\ref{fig:JV_intro} illustrates this effect by comparing two $JV$ curves -- one with transport losses and one without. The latter represents a hypothetical device where charge extraction occurs immediately after photogeneration. Similar to the external series resistance, the transport resistance causes a voltage drop ($\Vtr$) that reduces the slope of the $JV$ curve and lowers the fill factor. The term transport resistance losses refers to the effect of $\Vtr$ on the current--voltage characteristics. This voltage drop can be quantified by comparing the $JV$ curves at the same current density.\cite{wolf_series_1963} Unlike a constant series resistance, however, the transport resistance depends on conductivity, which varies with charge carrier concentration. Consequently, the conductivity decreases when moving along the JV curve from open-circuit to short-circuit. This makes transport-induced voltage losses more significant (see Figure~\ref{fig:vis} in section~\ref{sec:visual}). As a result, transport resistance influences the entire $JV$ curve -- impacting the slope near both open-circuit and zero applied voltage.\cite{wurfel_impact_2015,grabowski_fill_2022} 

In the absence of transport losses, the $JV$ curve only depends on recombination and its slope is inversely proportional to the recombination ideality factor ($\nid$). When transport losses are present, the slope decreases due to low conductivity. We parametrise this change through the apparent ideality factor ($\napp$) which accounts for both recombination and charge transport. The two $JV$ curves can be characterised by their fill factors: $\FF$ corresponding to the real solar cell that includes transport losses, and the pseudo-fill factor, $\pFF$, describing a hypothetical device where transport losses are absent. The difference between these fill factors represents the loss due to transport resistance. Rather than understanding this as a voltage loss, alternatively the fill factor difference can be seen as additional recombination caused by slow charge transport. Both viewpoints have merit, and independent of the perspective, understanding transport loss requires a thorough understanding of transport and recombination.

\subsection{Structure of the review}

After having briefly introduced the voltage drop due to a low conductivity of the active layer, in the following section~\ref{sec:meta-review} we want to highlight to the reader that even the best organic solar cells are typically limited by transport resistance losses. To demonstrate this, we performed a literature review on wide range of relatively recently published devices, for each device we estimated the fill factor losses due to the transport resistance. To give a qualitative impression of how transport resistance limits the fill factor, we visualise its effect in section~\ref{sec:visual}. The detailed theoretical background will be provided in section~\ref{sec:theory}, also paving the way to understand the later sections. The experimental determination of the transport resistance is described with instructions that can be easily followed (section~\ref{sec:exp}). Figures of merit can provide a quick way to determine whether a solar cell has low conductivity at a given light intensity and is therefore limited by transport resistance. We will discuss different figures of merit and what we can learn from each of them in section~\ref{sec:FoM}. It is very important to highlight that the voltage loss described up to here is one way to understand charge extraction losses, but there are other ways to express the same information: in section~\ref{sec:perspectives}, we present two additional facets of transport resistance losses, the \enquote{photoshunt} and the current loss due to additional recombination of slow charge carriers. Other factors in addition to recombination and transport resistance can also influence the fill factor, in section~\ref{sec:beyond} we will briefly highlight two of these effects; (i) electric field-dependent photogeneration (also known as geminate recombination); and (ii) recombination with injected charge carriers. Finally, we will discuss strategies to minimise fill factor loss due to transport resistance (section~\ref{sec:outlook}).

\section{The largest fill factor loss is due to the transport resistance.}\label{sec:meta-review}

\begin{figure*}[!tb]
    \centering
    \includegraphics[width=0.95\linewidth]{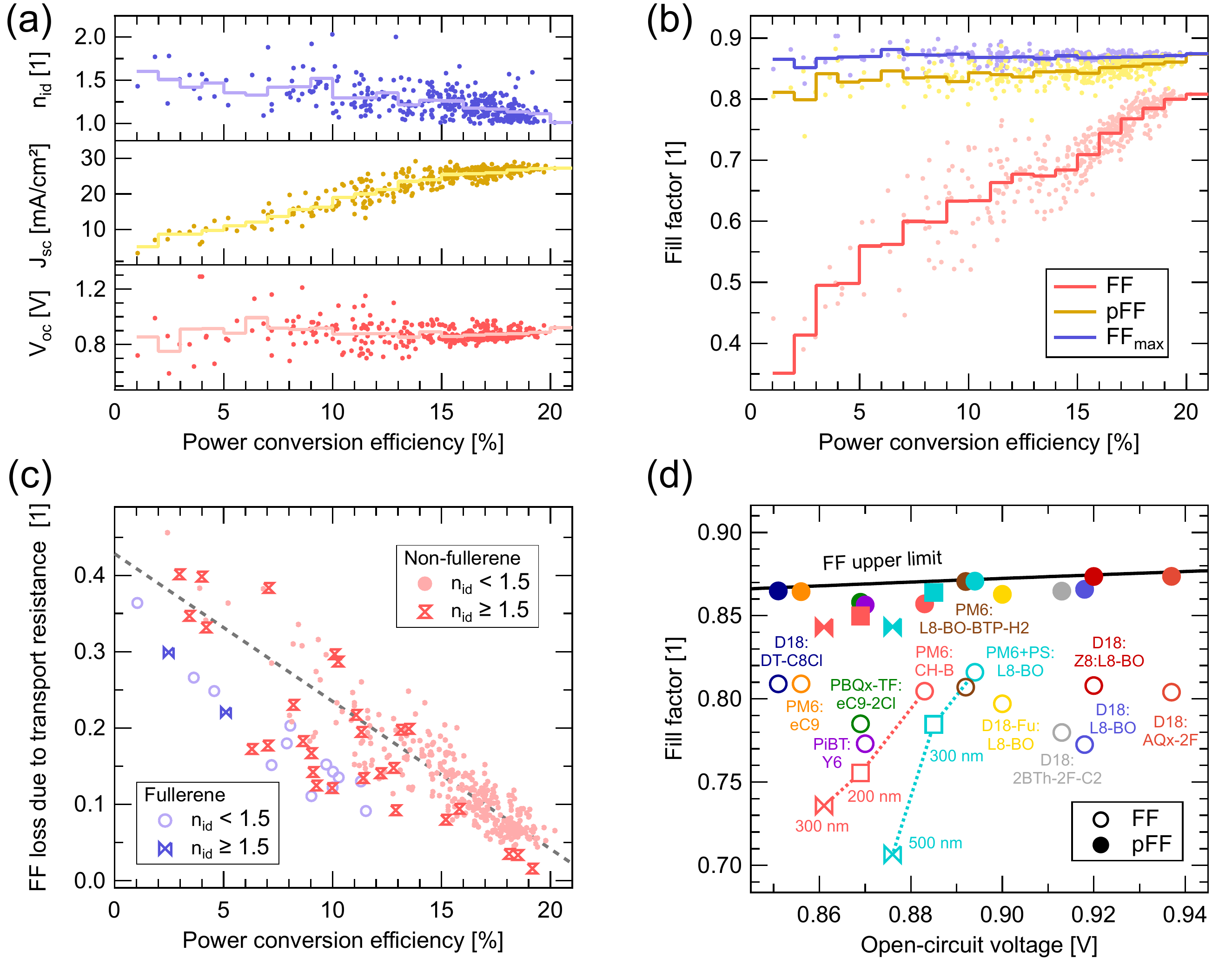}
    \caption{(a) Parameters $\nid$, $\jsc$, and $\Voc$ of various organic solar cells plotted as a function of their respective PCE. (b) Different types of fill factor versus PCE of solar cells, the cityscape line represents the binning of assembled data. (c) Estimated transport resistance loss versus PCE, the grey dashed line shows the correlation. (d) Comparison of $\pFF$ and $\FF$ in OSCs (PCE > $19\%$) based on different donor--acceptor blends. The $\FF$ upper limit (solid line) was determined using the Green equation~\labelcref{eqn:green} with $\nid = 1$.}
    \label{fig:review01}
\end{figure*}

In this section, we analyse how the transport resistance influences the $\FF$ and power conversion efficiency (PCE) of actual organic solar cells. We assembled data of the key parameters reflecting the transport and recombination in solar cells based on 125 publications, with 390 devices made from over 40 diverse photoactive material systems.\cite{liu202419EfficiencyBinary,qian_design_2018,arshad_hydrogen-bonding_2024,zhan2021LayerLayerProcesseda,jiang_non-fullerene_2024,cui2020SingleJunctionOrganicPhotovoltaica,fu_1931_2023,cai_wellmixed_2021,fan2019Achieving16Efficiencya,fan2018ChlorineSubstituted2Dconjugated,hong_ecocompatible_2019,zhao_environmentally_2018,yuan_fused_2019,sun2020HighEfficiencyPolymerb,song_high-efficiency_2021,yu_improved_2019,chen_modulating_2017,li_non-fullerene_2021,gao202016EfficiencyTernary,cui201916EfficiencyOrganica,xu_realizing_2018,cui2021SingleJunctionOrganicPhotovoltaica,zhang_single-layered_2021,fan_synergistic_2018,chang_synergy_2019,jin_thick_2017,zhang2020VolatilizableCosteffectiveQuinonebaseda,yu_wide_2017,li2023RefinedMolecularMicrostructurea,fu2024StackingModulationPolymera,bin_114_2016,dong_190_2024,chen_19_2024,zou_bithiazole-substituted_2023,pang_facile_2021,liang_rare_2023,zeng_achieving_2024,zeng_all-polymer_2023,gasparini_alternative_2015,wang_asymmetric_2022,yu_efficient_2024,pang_benzodthiazole_2023,wei_binary_2022,wang_binary_2023,chen_binary_2024,he_compromising_2022,li_conjugation-broken_2024,chen_delayed_2024,lee_efficient_2024,gan_electrostatic_2023,ye_enhanced_2024,song_film-formation_2023,zhao_fullerene-liquid-crystal-induced_2023,ma_high-efficiency_2023,gao_high-efficiency_2016,zhou_high-efficiency_2018,zhang_high-performance_2021,deng_high-performance_2024,qin_highly_2016,wan_highly_2017,oh_impact_2024,lin_isomerization_2024,chen_layer-by-layer_2022,he_manipulating_2022,lu_mechanistic_2015,zhan_multiphase_2022,zhan_over_2020,gao_over_2022,yuan_polythiophenes_2022,gu_precisely_2024,zhang_rational_2024,cao_rebuilding_2024,chen_simplified_2024,zhang_simultaneously_2024,sun_single-junction_2022,he_single-junction_2015,an_solution-processed_2020,zhu_suppressing_2024,duan_ternary_2022,feng_tuning_2022,tang_two_2020,li_unveiling_2021,cai_vertically_2022,sun_-extended_2024,chen_exploiting_2024,chen_restrained_2023,xu_volatile_2024,fan_enabling_2021,wang_high-performance_2022,yu_polymer_2024,li_revisiting_2022,xie_water-based_2024,wang_control_2023,liu_dimerized_2024,zhu_efficient_2022,zhang_fluid_2022,zhao_high-performance_2023,fan_high-performance_2022,sun_high-speed_2022,zhao_hot_2020,yue_meniscus-assisted_2022,xue_nonhalogenated_2022,yuan_patterned_2021,zhao_processing-friendly_2019,zhang_reducing_2022,liu_regulation_2023,cho_role_2023,lee_slot-and_2019,li_trialkylsilyl-thiophene-conjugated_2024,he_versatile_2022,wu_conjugated_2021,li_crystalline_2019,liu_efficient_2023,aubele_molecular_2022,li_narrow-bandgap_2021,yang_over_2023,cheng_regulating_2023,cheng_three--one_2024,he_unraveling_2022,pan_binary_2021,weng_high-efficiency_2020,gao_over_2020,wei_over_2024,wang_significantly_2020,wang_synergetic_2022} The year of publication ranges from 2015--2024, and the PCE varies from less than $5\%$ to over $20\%$, covering a few fullerene-based and many non-fullerene-based systems. We collected the reported solar cell parameters such as the open-circuit voltage ($\Voc$), the short-circuit current density ($\jsc$), $\FF$, PCE, and $\nid$ (from the suns-$\Voc$ measurements\cite{kerr_generalized_2004}), and determined estimates for the transport resistance loss using a theoretical framework described further below. Through this section's meta-review, we could gain insights into ways to reduce transport resistance losses that will be discussed in section~\ref{sec:outlook}.

The solar cell parameters $\Voc$, $\jsc$, and $\nid$ are shown in Figure~\ref{fig:review01}(a) with respect to the PCE of all assembled devices. $\Voc$ is concentrated in the range of 0.8--0.9~V for high-efficiency devices, which is likely determined by the energy level alignment of current state-of-the-art donor--acceptor combinations. The improvement in PCE was accompanied by a substantial increase in $\jsc$, due to the development of highly absorbing narrow bandgap non-fullerene acceptors. As a measure describing the recombination mechanism and, to some degree, the effective energetic disorder, the diode ideality factor $\nid$ decreases continuously towards unity as the PCE increases, indicating the highly suppressed energetic disorder in corresponding systems. 

One primary remaining parameter is the $\FF$, which is affected by several loss mechanisms, including singlet exciton losses by repopulation, geminate recombination, non-geminate recombination, and transport resistance losses.\cite{saladina_charge_2021,gohler_nongeminate_2018,woepke2022,saladina_transport_2024} 
The $\FF$ can be estimated using the empirical Green equation\cite{green1982accuracy}, 
\begin{equation}\begin{split}\label{eqn:green}
    \FF &= \frac{\voc-\ln(\voc+0.72)}{\voc+1},\quad \text{where} \\
    \voc &= \frac{e\Voc}{\napp\kT} , 
\end{split}\end{equation}
with the apparent ideality factor, $\napp$, contained in the normalised open-circuit voltage. The Green equation was confirmed to hold for a wide range of organic solar cells devices under different measurement conditions, when the ideality factor is adapted to account for the transport resistance.\cite{neher2016new,saladina_transport_2024} Furthermore, \cref{eqn:green} provides an upper limit of the fill factor ($\FFmax$) that corresponds to the detailed balance efficiency limit,\cite{shockley_detailed_1961} assuming no leakage currents, no series resistance and a unity ideality factor ($\napp=1$). When instead entering the experimentally determined $\nid$ into the Green equation, we can predict the $\pFF$: the fill factor that the device would have, were it not transport limited. 

\begin{figure*}[!tb]
    \centering
    \includegraphics[trim = 0cm 0cm 0cm 1cm, width=1\linewidth]{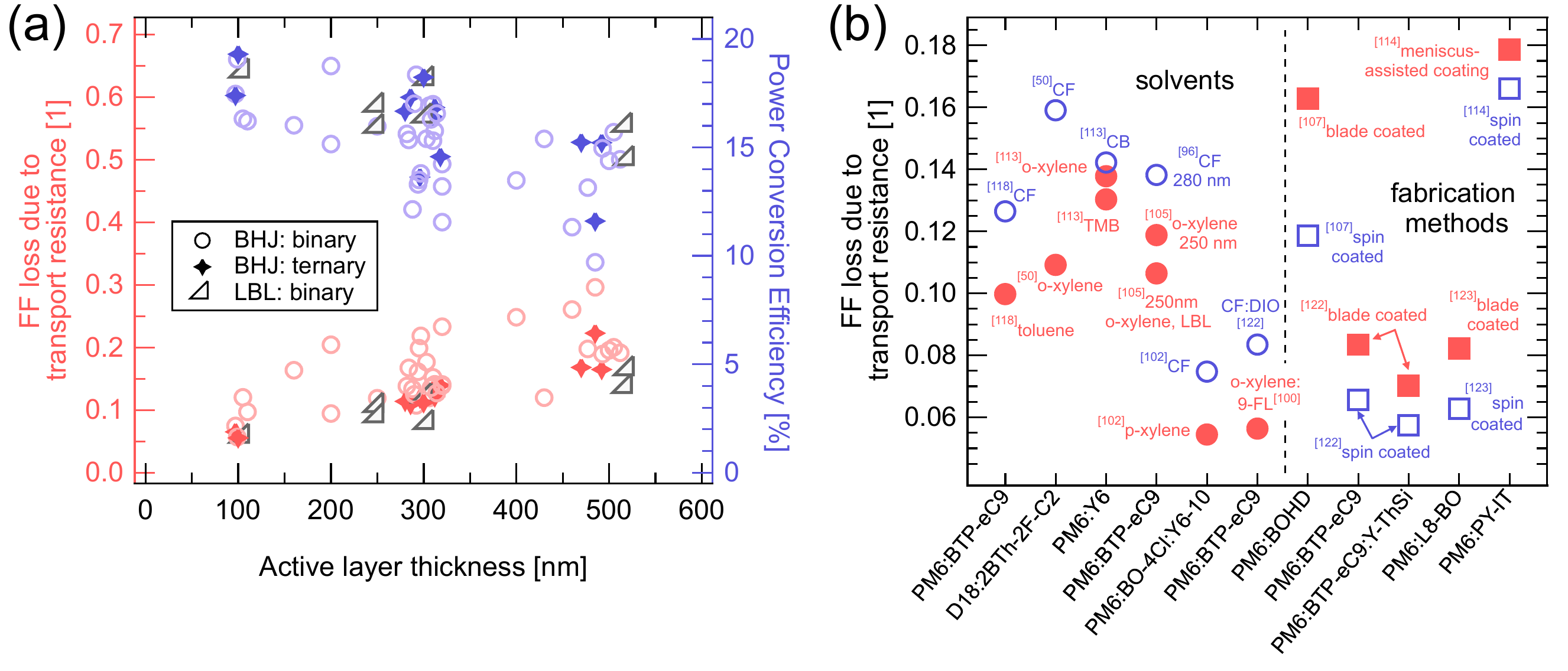}
    \caption{$\FF$ loss due to transport resistance for solar cells grouped by: (a) active layer thickness, with devices using binary (circles) and ternary (stars) BHJ or LBL-deposited (triangles) active layers. PCE is indicated for reference; (b) active layers processed from halogenated vs.\ non-halogenated solvents (left of the dashed line), and spin-coated vs.\ non-spin-coated active layers (right of the dashed line).}
    \label{fig:review02}
\end{figure*}

In Figure~\ref{fig:review01}(b), the $\pFF$ approaches $\FFmax$ as the PCE increases. This shows that the geminate and nongeminate recombination, previously considered primary contributors to $\FF$ loss, are less limiting in state-of-the-art OSCs. This can be attributed to the development of novel non-fullerene acceptors, whose favourable packing properties effectively reduce the degree of energetic disorder in the active layer.\cite{wang_physical_2024} On the other hand, while the device $\FF$ has been increasing with the PCE, there is still a significant gap between $\FF$ and $\pFF$. The difference between the actual $\FF$ value and the $\pFF$ originates mostly from the transport resistance, although this fact does not seem well-known in the OPV research community. 

Figure~\ref{fig:review01}(c) presents the fill factor loss due to transport resistance, determined as $\pFF-\FF$, versus PCE. Lower transport resistance losses yield higher PCE values. By setting a filter for fullerene acceptor-based systems and $\nid$ ($\geq$1.5, significant energetic disorder), we can effectively filter most outliers from the plots. This clearly shows that transport resistance dominates the $\FF$ and PCE loss in the vast majority of non-fullerene-based OSC devices, despite the fact that this is usually not mentioned in the corresponding references. In Figure~\ref{fig:review01}(d) we highlighted that even the most efficient systems with PCE $>19\%$ exhibit significant $\FF$ loss due to transport resistance. In an ideal system without this loss, the PCE of D18:Z8:L8-BO -- with a certified record of $20.2\%$ -- could be improved to $22\%$. Moreover, the transport resistance loss increases rapidly as the photoactive layer thickness increases (PM6:CH-B and PM6+PS:L8-BO serve as examples). This further emphasises that transport resistance is a critical challenge for commercial solar cells.

Lab-scale organic solar cells can be precisely fabricated with 100~nm photoactive layers by spin-coating from solution. However, for large-area roll-to-roll printing, thicker active layers of several hundred nanometres may be desirable for getting pin-hole-free films at a high printing rate.\cite{park_progress_2020} \cref{eq:Rtr} indicates that transport resistance loss is a strong function of active layer thickness. Analysing the collected data from literature confirms this trend, as shown in Figure~\ref{fig:review02}(a). The transport resistance loss rises by two to three times when the active layer thickness is increased from 100~nm to 500~nm. In the figure, we highlighted devices consisting of ternary blends or made by layer-by-layer (LBL) deposition. Both represent feasible strategies to lower the transport resistance losses at higher active layer thicknesses.

Another long-term goal in large-scale OSCs fabrication is eliminating halogenated solvents and resorting to green alternatives. While spin coating is incompatible with large-scale manufacturing, the literature suggests that employing non-halogenated solvents or non-spin-coating methods decreases the device efficiency.\cite{schrodner_reel--reel_2012} In Figure~\ref{fig:review02}(b) we analysed the transport resistance losses across multiple high-efficiency non-fullerene blend systems. The active layers processed with non-halogenated solvents achieve lower transport resistance losses for the same active layer thickness than those processed with halogenated ones, indicating that processing with non-halogenated solvents is well-optimised. This paves the way for adopting non-toxic and environmentally friendly active layer inks for mass production. In contrast, the non-spin-coating methods still tend to result in higher transport resistance losses in various donor--acceptor systems at the current stage of the research community. Finding and optimising the most suitable coating techniques to fabricate high-quality photoactive layers with comparably optimised morphologies remains challenging.

We evaluated subsets of the collected data from specific aspects, such as device configurations and photoactive layer additives. Regarding device configuration, we note that high-efficiency cells preferentially adopt the conventional structure, probably due to the photocatalytic effect between the ZnO and non-fullerene-acceptors, which has been largely overcome by viable strategies proposed by the research community.\cite{jiang_photocatalytic_2019,li_achieving_2022} In the data we assembled, the adoption of an inverted device structure lags far behind the conventional structure (71 devices vs.\ 319 devices). The lower degree of optimisation of the corresponding photoactive layers leads, on average, to higher transport resistance losses. Nevertheless, employing ultra-thin electron-transport layers, low work function top electrodes, and corrosive acidic PEDOT:PSS hole-transport layers in conventional devices pose significant challenges for robust mass production.\cite{cheng_stability_2016} Future industrially solution-processable large-area OPV modules rely on the inverted structure.\cite{chang_unveiling_2024} Therefore, we encourage the research community to focus more on developing inverted devices to reduce the transport resistance loss gap between two device configurations. Using this dataset we also found that the photoactive layer using 1,8-diiodooctane (DIO) shows slightly lower transport resistance loss, probably due to better fine-tuning of the nanomorphology. More importantly, the absence of additives does not appear to result in higher transport resistance loss compared to using DIO, 1-chloronaphthalene (1-CN), or even more advanced solid additives in non-fullerene solar cells. A roll-to-roll process design should inherently avoid using additives due to issues such as solvent residues -- potential contaminants undermining the device stability -- and increased process complexity. This seems to be more achievable in non-fullerene systems.

We established that the transport resistance is a loss mechanism that is limiting all state-of-the-art organic solar cells. The first step towards minimising its effects is gaining a deeper understanding of what transport resistance is.

\section{Visualisation of transport losses}\label{sec:visual}

\begin{figure*}[!tb]
    \centering
    \includegraphics[width=0.9\textwidth]{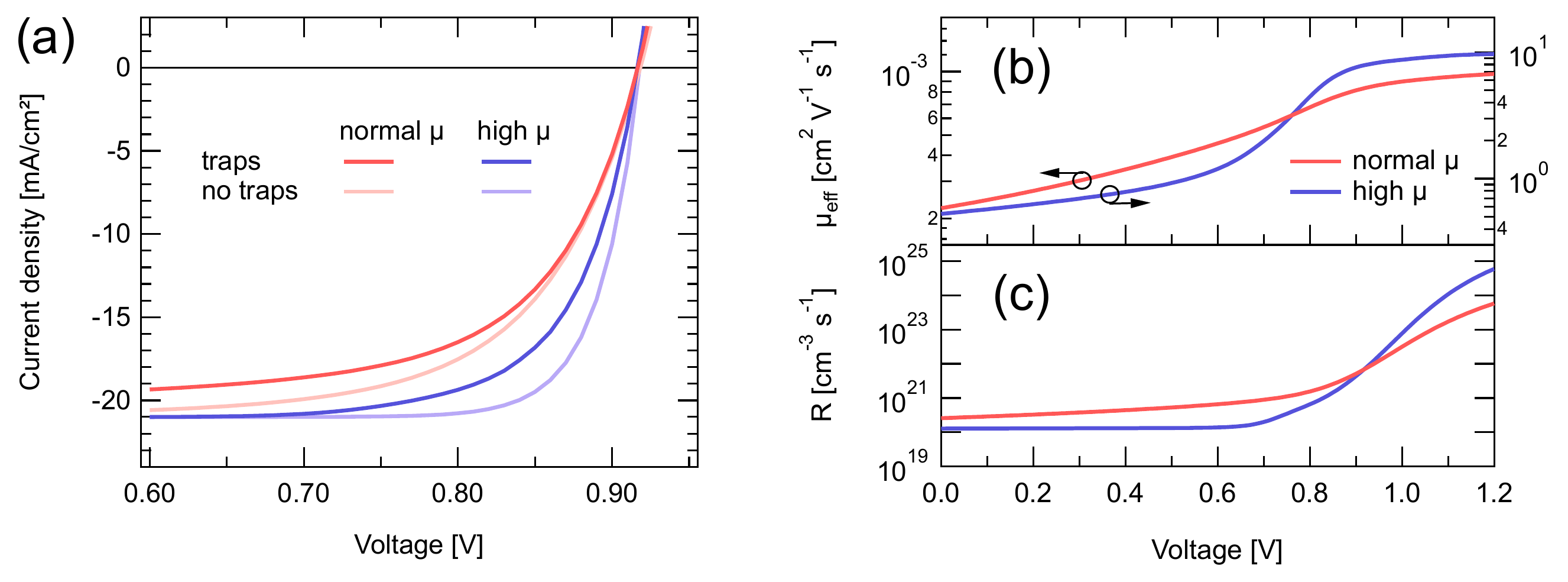}
    \caption{Drift--diffusion simulation based on a 200~nm thick PM6:Y6 solar cell.\cite{woepke2022} 
    (a) Current--voltage characteristics for the cases with and without energetic disorder (denoted as traps and no traps, respectively), for the case of high (violet line) and normal mobility (red line). Transport-resistance-free curves (not shown) virtually coincide with the high mobility curves for both cases. The values of the charge carrier mobilities are given in (b), and the recombination rate is plotted in (c) as a function of applied voltage at 1~sun. Mobility is a strong function of applied voltage. The device with a higher mobility has a lower charge density, and hence, lower recombination rate, due to better carrier extraction; however as $\Voc$ is approached recombination equals generation and the recombination rates become equal.} 
    \label{fig:vis}
\end{figure*}

Transport resistance is a loss that affects the fill factor and the short-circuit current density, and therefore the power conversion efficiency. It is due to a non-ideal charge extraction caused by a low effective active layer conductivity. In this section, we simulate current--voltage curves using realistic material parameters, including energetic disorder that is found in organic solar cells. Our main aim is to use these simulations to understand and visualise the impact of both transport resistance on device behaviour and its relationship with disorder.

In \cref{fig:vis}(a), the $JV$ curve of an OPV device with an active layer thickness of 200~nm is shown, illuminated with a light intensity equivalent to 1~sun that corresponds to $100\,\mathrm{mW(cm)^{-2}}$ under a AM1.5G conditions. 
The simulation parameters are based on simultaneously fitting multiple experiments under different conditions (different light intensities, at room temperature).\cite{woepke2022} The parameters were adapted to contain perfectly selective contacts to avoid an impact of surface recombination on the simulation results. This leads to an open-circuit voltage that is slightly higher than expected. The dark red line represents the real OPV device, in which the energetic disorder (traps) leads to a typical, voltage-dependent mobility (\enquote{normal $\mu$}) that is shown in \cref{fig:vis}(b) and ranges roughly from $2\E{-4}$ to $1\E{-3}\,\mathrm{cm^2(Vs)^{-1}}$. As this is a thicker-than-normal device, the fill factor loss is pronounced. For comparison, the $JV$ curve with the same parameters but a roughly three orders-of-magnitude higher effective charge carrier mobility (\enquote{high $\mu$}, violet line) has a much higher fill factor and even a higher short-circuit current density. The recombination rates are independent of the charge carrier mobility and, therefore, the recombination rates presented in \cref{fig:vis}(c) are equal at $\Voc$. There, due to $J=0$, the transport resistance loss is zero. In the high mobility case, the fill factor is only determined by recombination (with a voltage dependence represented by the ideality factor $\nid$), but the charge extraction is (close to) perfect. In the normal mobility case, in addition to recombination, the $JV$ curve is limited by the transport resistance. As the transport resistance is due to a low active layer mobility that limits the charge collection, the fill factor and, thus, the PCE, could also be significantly higher if the charge collection was higher. 

While in most of this Perspective, we describe the transport resistance as a voltage loss -- see for instance the theoretical framework in section~\ref{sec:theory} -- the carrier concentration (and therefore voltage) dependent recombination rate shown in \cref{fig:vis}(c) for $V < \Voc$ can also be seen in view of a current loss at the same voltage due to the transport resistance (section~\ref{sec:currentloss}). 

It is important to note that the suns-$\Voc$ curve of the normal mobility case (not shown), shifted down by the generation current under 1~sun, coincides completely with the high mobility case: as the open-circuit voltage is measured at zero current, the transport resistance is also zero in this particular case. This is why the suns-$\Voc$ method is an excellent tool to predict the $JV$ characteristics if the solar cell had perfect charge extraction. The comparison to the transport-limited illuminated $JV$ curve allows to quantify the transport losses in real organic solar cells, as described in section~\ref{sec:exp}.

The light red and violet lines in \cref{fig:vis}(a) represent the case where energetic disorder, or traps, have been turned off in the simulation. This forces all charge carriers to be mobile, and results in an increased average mobility over all charge carriers. Furthermore, the recombination rate will also change as free-to-trapped recombination is not possible. Therefore, to make the simulations more comparable to the realistic case with traps, the free-to-free carrier recombination prefactor was adjusted so to make the $\Voc$s identical. In addition, the mobility of mobile charge carriers was set to the average charge carrier mobility of the device with traps at the maximum power point. On one hand, it becomes evident that despite these measures, the $JV$ curves without traps show a strong impact of the transport resistance. The reason is that the low charge carrier mobility (in the normal mobility case) determines the effect and not traps as such. On the other hand, without traps the recombination ideality factor $\nid$ is unity (similar to a case with a Gaussian density of states) and the general shape of the $JV$ curves differs from the realistic case with traps. This underlines the importance of taking energetic disorder into account when trying to understand transport resistance losses. 

So what happens to the current density of the realistic solar cell with traps (normal mobility, \cref{fig:vis}(a), dark red line) when the voltage is changed from short circuit towards open circuit? At short circuit, most photogenerated charge carriers are able to leave the device, resulting in a relatively low carrier density. However, due to the transport resistance and the fact that not all carriers are mobile, the short-circuit current density remains lower than the generation current density. As $\Voc$ is approached, carrier extraction slows and the charge carrier density increases. The charge carriers can occupy higher energy trap states, resulting in more mobile charge carriers, and the charge carrier mobility increases (\cref{fig:vis}(b)). Transport resistance in an organic solar cell is therefore a function of the density of trap states -- or the degree of energetic disorder -- and their occupation. A lower density of trap states is beneficial to minimise transport losses, as it would increase overall mobility and reduce the transport resistance by forcing more charge carriers to be mobile and less to be trapped. 

After having gained a feeling for how transport resistance affects $JV$ curves, we will provide the detailed theoretical background in the following section.

\section{What is transport resistance and why is it a loss?}\label{sec:theory}

The current--voltage measurement of a solar cell under illumination is the simplest and most central characterisation technique for photovoltaic devices as it determines the PCE and, therefore, functionality of the device. Understanding efficiency limiting mechanisms in photovoltaics is therefore intricately linked to investigating the influence of material properties on the current--voltage curve. The origin of the current--voltage curve is the continuity equation, which considers the conservation of charge carriers. The charge carrier density can increase via photogeneration quantified via the generation rate, $G$, it can decrease by recombination (quantified by the recombination rate, $R$) and it can increase or decrease if there is a change -- more specifically, a divergence -- of the current density $J$ at any given point in space.\cite{sze2007physics-book} For steady-state conditions, with time-independent carrier concentration $n$, the continuity equation for one charge carrier type is given by 
\begin{equation}\begin{split}\label{eq:continuity}
    0 &= G - R + \frac{1}{e}\cdot \nabla J , 
\end{split}\end{equation}
with $e$ denoting the elementary charge. 

To determine the current density, one integrates the above equation over the sample thickness $d$, which leads to 
\begin{equation}\begin{split}\label{eq:opv-diode}
    J 
    &= ed (\bar{R} - \bar{G}) \\
    &= J_0 \cdot \exp\bl \frac{\EF}{\nid\kT}\br - \jgen . 
\end{split}\end{equation}
Here, $\bar{R}$ and $\bar{G}$ denote the spatially averaged recombination and generation rates, $\EF$ stands for the average quasi-Fermi level (QFL) splitting, $\nid$ the recombination ideality factor, $k_\mathrm{B}$ the Boltzmann constant, and $T$ the temperature. The generation current density $\jgen$ accounts for both photogeneration under illumination, and thermal generation in the dark, the latter being equal to $J_0$. 

The relation between $\bar{R}$ and the $\EF$ in \cref{eq:opv-diode} can be established by first expressing the $\bar{R}$ in terms of the charge carrier density $n$, i.e., $\bar{R} \propto n^\delta$, and then linking the charge carrier density to the average QFL splitting via $n=n_0\exp\bl \EF/\nn\kT \br$. Here, $n_0$ is the dark carrier density at zero $\Delta E_F$. In organic solar cells, the parameter $\nn$ and the recombination order $\delta$ reflect how the charge carrier dynamics are influenced by the density of states. Together they shape the ideality factor $\nid$ in \cref{eq:opv-diode}, determining how the recombination rate relates to the QFL splitting and temperature.\cite{kirchartz2012meaning,hofacker2017dispersive,saladina_power-law_2023} 

The key downside of \cref{eq:opv-diode} is that it expresses the current density as a function of $\EF$ but not as a function of a measurable voltage. Thus, the traditional approach in photovoltaics would be to define a diode equation under illumination, where the current density is expressed in terms of the \emph{externally applied voltage}, $\Vext$.\cite{shockley1949theory,wurfel2016physics-book} As the externally applied voltage differs from the average $\EF/e$ at least by a voltage drop over ohmic resistances in series to the diode (e.g., caused by contact layers), the current--voltage curve of an illuminated diode is typically written as
\begin{equation}\begin{split}\label{eq:opv-diode-extended}
    J &= J_0 \cdot \exp\bl \frac{e(\Vext - J\Rs)}{\nid\kT} \br\\
    &\qquad + \frac{\Vext - J\Rs}{\Rp} - \jgen . 
\end{split}\end{equation}
Here, $\Rs$ denotes the series and $\Rp$ the parallel resistance of the solar cell. In the traditional interpretation of the equation, $\Rs$ represents any ohmic resistances of the diode that could originate either from contact layers, or from doped regions of the absorber layer of the solar cell. For instance, in a silicon solar cell based on a p-type wafer, the hole transport along the thickness of the p-type wafer should lead to an ohmic resistance. The conductivity of the p-type wafer is fixed by the doping density and does not depend on illumination as long as the photogenerated carrier density does not exceed the doping density. However, in case of organic solar cells, where the resistance of the semiconducting layers of the device originates primarily from space charge regions within the absorber layer, the internal series resistance becomes a strong function of the mobile carrier density (rather than the doping density). As the carrier concentration depends exponentially on the voltage, this will immediately lead to a non-ohmic resistive element. Furthermore, the effect of illumination on the carrier density is coupled to the effect of the applied voltage by the conductivities of the involved semiconducting layers. Thus, especially for low-conductivity materials such as molecular semiconductors, the effect of illumination on the effective resistance of the absorber layer is significant. Hence, to fully align the classical diode model with the reality of \cref{eq:opv-diode} in low-mobility semiconductors, it is necessary to interpret the series resistance in a more general way. Rather than being a purely ohmic resistance, the parameter becomes an effective resistive term that depends non-linearly on both voltage and light intensity. To acknowledge this change in meaning, this resistance is referred to as transport resistance.

\subsection{Transport resistance: connecting the continuity and diode models}\label{sec:theory-vtr}

The difference between the models based on the continuity and diode equations, as previously mentioned, lies in the type of voltage they consider. To enable a better comparison between $\EF$ and the external voltage, it is convenient to introduce the so-called \emph{implied voltage} as a way to express $\EF$ in units of a voltage via 
\begin{equation}
    \Vimp = \frac{\EF}{e}. 
\end{equation} 
Only in the limit of infinite conductivities of the absorber layer of a solar cell, the implied voltage equals the external voltage $\Vext=\Vimp$.\cite{sze2007physics-book} In practice, this is never true at short circuit in any type of solar cell, but the differences become larger, the lower the mobilities and conductivities of the absorber layers are. Thus, the discrepancy is particularly significant in organic solar cells with their slow charge carrier transport.\cite{wurfel_impact_2015} 

\begin{figure*}[!tb]
    \centering
    \includegraphics[width=1\textwidth]{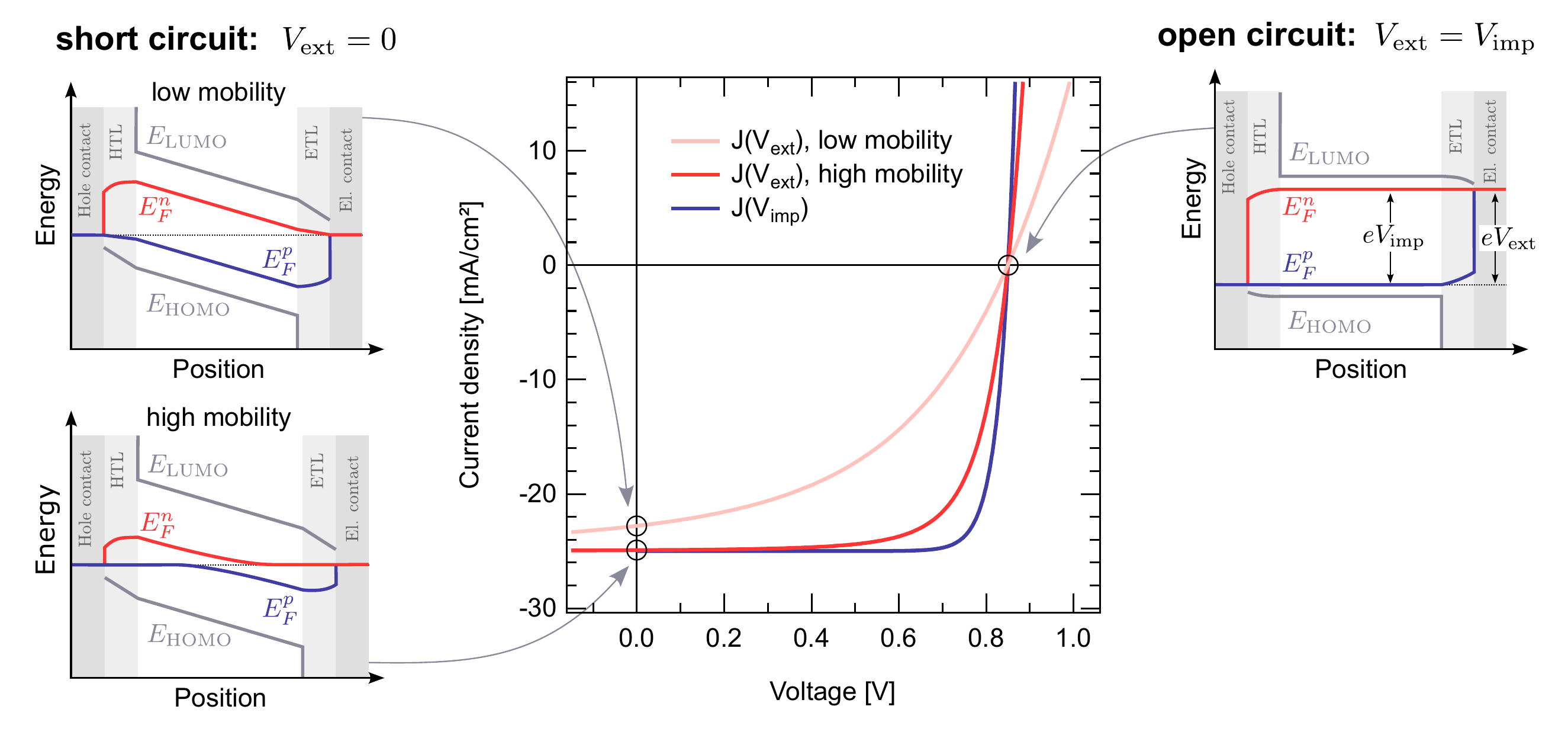}
    \caption{$JV$ curves of solar cells with high and low mobility, calculated using \cref{eq:j_Vext}. The schematic energy-level diagrams show that: at open-circuit, $\Vext=\Vimp$, and the QFLs are flat; at short-circuit, low mobility leads to a larger QFL gradient and difference between $\Vext$ and $\Vimp$, compared to the high mobility case. The $J(\Vimp)$ represents a solar cell without transport resistance (infinite conductivity) and flat QFLs, where $\Vext=\Vimp$ at any point of the curve. After Ref.~\parencite{wurfel_impact_2015}.}
    \label{fig:theory}
\end{figure*}

The link between the two descriptions, Equations~\eqref{eq:opv-diode} and \eqref{eq:opv-diode-extended}, is established through the concept of transport resistance -- the resistance to the flow of charge carriers within the solar cell. Its effect on the $JV$ curve is similar to the series resistance of the circuit -- it causes a voltage drop 
\begin{equation}\label{eq:Rtr}
    \Vtr = J \Rtr = J \frac{d}{\sigma} . 
\end{equation}
Here, $\Rtr$ is the transport resistance, and $\sigma$ is the effective conductivity, which depends on the conductivity of both electrons and holes as described in section~\ref{sec:FoM}. In the following, we demonstrate how transport resistance serves as a bridge between the models based on the continuity and the diode equations. 

To understand transport resistance, we start with the relationship between current density, conductivity, and the QFL gradient $\nabla E_F$:
\begin{equation}\label{eq:j_sigma}
    J = \frac{\sigma}{e}\cdot \nabla E_F . 
\end{equation}
The derivation using the Boltzmann transport equation can be found in Ref.~\parencite{nelson2003physics-book}. This expression accounts for all non-equilibrium effects on charge transport, encompassing both drift and diffusion.\cite{sze2007physics-book} 
The gradients in QFLs represent a general principle driving charge transport. The drift--diffusion model emerges as a specific case of this broader concept, where the QFL gradients translate into drift and diffusion currents through the material. 

The impact of the QFL gradient on the difference between the implied and external voltage can be understood through the band diagrams in \cref{fig:theory}. 
Under open-circuit conditions, when the QFLs are flat, the resulting current density is zero. The externally applied and implied voltages are equal. 
At short circuit, even though $\Vext=0$, charge carriers take time to be extracted, hence the implied voltage corresponding to the QFL splitting is non-zero and still at a higher value below $\Voc$. Lower charge carrier mobility results in lower conductivity and a higher QFL gradient, leading to a larger mismatch between $\Vext$ and $\Vimp$. The gradient is mainly caused by the active layer of the solar cell, with little contribution from the typically highly conductive transport layers. 

The \cref{fig:theory} shows that under the assumption of constant and equal QFL gradients, the difference between the applied and implied voltages is given by $d \cdot \nabla E_F/e$. Comparison of Equations~\labelcref{eq:Rtr} and \labelcref{eq:j_sigma} reveals that this difference is equal to the voltage drop $\Vtr$ that is caused by the transport resistance,\cite{schiefer_determination_2014,neher2016new}
\begin{equation}\label{eq:Vtr-deltaV}
    \Vtr = \Vext - \Vimp = d\cdot \frac{\nabla E_F}{e} . 
\end{equation}
Having outlined the concept of transport resistance, we move to the description of its impact on the current--voltage characteristics. 
We will compare two cases: an ideal solar cell without transport resistance losses, and a real solar cell, where $\Vext$ and $\Vimp$ are not equal. 

In a solar cell without transport resistance losses, denoted $J(\Vimp)$ in \cref{fig:theory}, the implied and external voltages are equal even at a non-zero current flow. This equality is achievable only when the QFLs are flat, which, based on \cref{eq:j_sigma}, occurs either when $J=0$ (at $\Voc$) or when the conductivity is infinitely high. Thus, an ideal solar cell effectively has infinite conductivity. For this type of solar cell, the diode equation is expressed via \cref{eq:opv-diode}, where $\Vimp=\EF/e$ can be replaced by the applied voltage $\Vext$. 

In real solar cells, conductivity is not infinite, hence there is always a discrepancy between $\Vext$ and $\Vimp$ (except at $\Voc$). For such cases, \cref{eq:opv-diode} must be adjusted to account for the impact of transport resistance while remaining expressed in terms of the measurable parameter $\Vext$. By using \cref{eq:Vtr-deltaV} to relate the implied and external voltages, we obtain \cite{wurfel_impact_2015}
\begin{equation}\begin{split}\label{eq:j_Vext}
    J &= J_0 \cdot \exp\bl \frac{e\bl\Vext - \Vtr\br}{\nid \kT} \br - \jgen .
\end{split}\end{equation}
Note that for clarity, at this stage the analysis omits the effects of parasitic resistances, as defined in \cref{eq:opv-diode-extended}, and only includes the effect of low conductivity via transport resistance. 

Equations~\eqref{eq:opv-diode} and \eqref{eq:j_Vext} are equivalent when evaluated at the same current density, since the applied and implied voltages are linked via the voltage drop $\Vtr$. 
To unify the format of these expressions, we introduce an apparent ideality factor allowing for transport resistance losses to appear implicitly in the denominator:\cite{saladina_transport_2024} 
\begin{equation}\label{eq:napp}
    \napp 
    = \nid\bl 1 + \frac{\Vtr}{\Vimp-\Voc} \br . 
\end{equation}
Additionally, we make use of the convenient property that at $J=0$, the recombination and generation currents are balanced. This allows us to express the generation current in terms of $\Voc$: 
\begin{equation}\label{eq:jgen}
    \jgen = J_0\exp\bl \frac{e\Voc}{\nid\kT} \br , 
\end{equation}
thereby replacing $J_0$ with $\jgen$ before the exponential term.\cite{wurfel_impact_2015,neher2016new} Using this substitution, current density can be consistently expressed in terms of either applied or implied voltage as:
\begin{equation}\begin{split}\label{eq:j_Vext_napp}
    J
    &= \jgen \left[ \exp\bl \frac{e\bl\Vimp - \Voc\br}{\nid \kT} \br - 1 \right] \\
    &= \jgen \left[ \exp\bl \frac{e\bl\Vext - \Voc\br}{\napp \kT} \br - 1 \right] . 
\end{split}\end{equation} 
This form of the diode equation is more suitable than the original one since the transport resistance stretches the $JV$ curve relative to $\Voc$ (we note that in the original diode equation, this stretching is given implicitly by $J_0$ and $\jgen$, but not explicitly). This approach allows us to directly compare the $JV$ curves for cases with and without transport resistance by examining the ideality factors and the slopes of these curves. 

The ideality factors in the above expression capture distinct device limitations -- $\nid$ reflects the impact of recombination, while $\napp$ accounts for both recombination and transport resistance. Under open-circuit conditions, the apparent ideality factor is expressed as $\napp = \nid+\alpha$, with\cite{neher2016new,saladina_transport_2024}
\begin{equation}\begin{split}\label{eq:alpha-general}
    \alpha &= \frac{ed}{\kT} \cdot \frac{\jgen}{\sigmaoc} . 
\end{split}\end{equation}
Here, $\sigmaoc$ represents the effective conductivity at $\Voc$, which depends on the conductivities of both electrons and holes. 
The parameter $\alpha$ reduces the slope of the $J(\Vext)$ curve relative to the scenario without transport resistance, serving as a figure of merit for quantifying transport resistance losses. It is governed by the interplay between recombination (noting that $\jgen=\jrec$ at $\Voc$) and conductivity, with their interaction driving fill factor losses rather than conductivity alone. We will consider $\alpha$ and further figures of merit in Section~\ref{sec:FoM}.

\subsection{Impact of transport resistance on the current--voltage curve slope}\label{sec:theory-slopes}

In solar cells based on high-mobility materials, transport resistance is negligible, allowing the total series resistance to be approximated by the constant external series resistance. Under this assumption, the slope of the $JV$ curve provides information about the ohmic resistances: the external series resistance is derived from the slope near $\Voc$, while the parallel resistance is obtained from the slope near 0~V. In low-mobility organic solar cells, the slope of the $JV$ curve is influenced by the transport resistance both near $\Voc$ and 0~V, leading to changes in the apparent series and parallel resistance values. 

To illustrate how transport resistance affects the slope of the $JV$ curve, we will exclude the impact of ohmic resistances for simplicity. Then, the slope can generally be determined as
\begin{equation}\begin{split}\label{eq:JV-slope-short}
    \frac{\der \Vext}{\der J} 
    &= \frac{\der\Vimp}{\der J} + \Rtr + J\frac{\der \Rtr}{\der J} . 
\end{split}\end{equation} 
Since the transport resistance depends on the current density, the slope is influenced not only by the value of $\Rtr$, but also by its derivative. The former dominates the slope around $\Voc$, where current density is small, while the latter becomes more significant at higher current densities, closer to short-circuit. 

To evaluate the derivative, we express $\Rtr = d/\sigma$ in terms of current density. Since conductivity depends on charge carrier density, it can be linked to the QFL splitting, similar to the relationship between recombination current and QFL splitting in \cref{eq:opv-diode}. Adopting a transport ideality factor $\nsig$, conductivity is expressed as 
\begin{equation}\label{eq:sigma}
    \sigma = \sigma_0 \exp\bl \frac{e\Vimp}{\nsig\kT}\br , 
\end{equation}
where $\sigma_0$ denotes the effective conductivity in the dark when the QFL splitting is zero. This allows to relate $\sigma$ to $\jrec=J+\jgen$ through the ideality factors $\nid$ and $\nsig$, and then differentiate it with respect to $J$. We get a general expression for the slope of the $JV$ curve at any point:
\begin{equation}\begin{split}\label{eq:JV-slope-general}
    \frac{\der \Vext}{\der J} 
    &= \frac{\nid\kT}{e\jrec} + \Rtr\bl 1 -\frac{\nid}{\nsig}\cdot\frac{J}{\jrec}\br . 
\end{split}\end{equation} 

Under open-circuit conditions, where no current flows, the term with $J$ (coming from the derivative term of $\Rtr$) vanishes. This allows to directly determine the effective conductivity from the difference in slopes of $\Vext(J)$ and $\Vimp(J)$:\cite{saladina_transport_2024}
\begin{equation}\begin{split}\label{eq:JV-slope-sigma}
    \frac{d}{\sigmaoc} &= \left.\frac{\der \Vtr(J)}{\der J}\right|_{J=0} . 
\end{split}\end{equation}
At $\Voc$, the slope of $\Vext(J)$ exceeds that of $\Vimp(J)$ by a term inversely proportional to conductivity, leading to an apparent series resistance even if the external series resistance is negligible. After factoring out $\jgen$, the slope at open circuit can be expressed as 
\begin{equation}\begin{split}\label{eq:JV-slope-voc}
    \left.\frac{\der\Vext}{\der J}\right|_{J=0}
    &= \frac{\kT}{e\jgen}\left( \nid + \alpha \right) . 
\end{split}\end{equation}
It is important to emphasise that that conductivity alone is insufficient to predict the fill factor; instead, the figure of merit $\alpha$ links the fill factor to the $JV$ curve slope. In this context, the slope must be scaled by the generation current density $\jgen$ to predict the fill factor. 

The slope of the current--voltage curve is also influenced by the transport resistance at zero applied voltage. This effect is referred to as the \enquote{photoshunt} -- the light intensity-dependent component of the apparent parallel resistance.\cite{wurfel_impact_2015,grabowski_fill_2022} 
Under short-circuit conditions, the first term in \cref{eq:JV-slope-general} becomes negligible, and the slope is dominated by transport resistance and its derivative. It can be approximated as
\begin{equation}\begin{split}\label{eq:JV-slope-jsc}
    \left.\frac{\der \Vext}{\der J}\right|_{\Vext=0} &\approx 
    \frac{d}{\sigmasc} \cdot \frac{\nid}{\nsig} \cdot \frac{\jgen}{\jgen-\jsc} , 
\end{split}\end{equation} 
where $\sigmasc$ is the effective conductivity at short-circuit, and $\jsc = -J$ at 0~V. Once again, we observe that the slope is inversely proportional to $\sigma$. Although parallel resistance is not explicitly included in this analysis, transport resistance modifies the $JV$ curve slope, resulting in the apparent photoshunt. We will further explore this concept in section~\ref{sec:Rphoto}, and examine its relationship to the fill factor and the figure of merit $\alpha$. Notably, to connect the slope at 0~V to the fill factor, it must again be scaled by $\jgen$, just as it was for the open-circuit case. 

Prepared with this theoretical background on the transport resistance, we move to the description of how the latter is determined experimentally.

\section{How to determine the transport resistance?}\label{sec:exp}

\begin{figure*}[!tb]
    \centering    \includegraphics[width=0.95\textwidth]{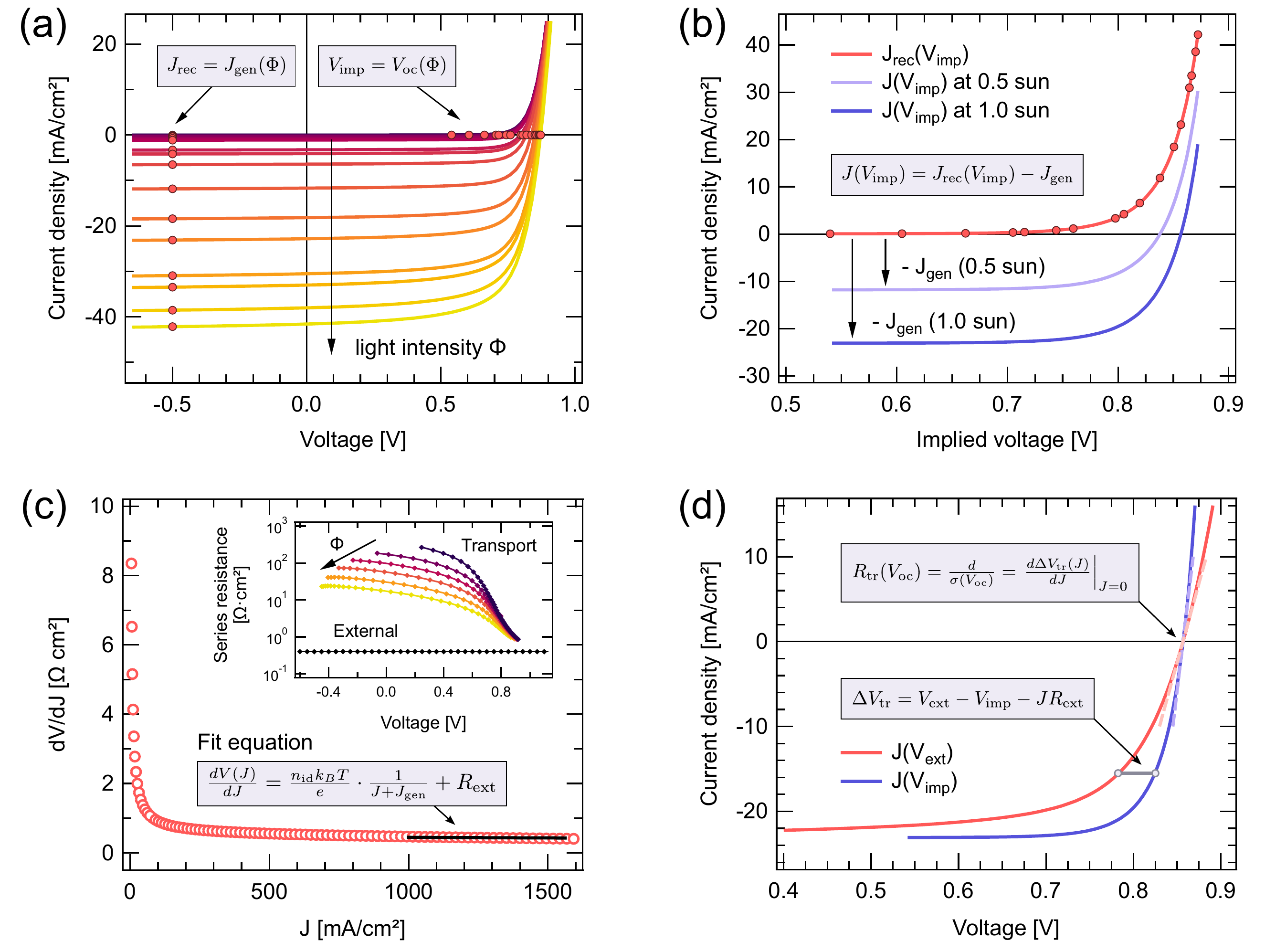}
    \caption{Experimental data for transport resistance evaluation with details given in the text. (a) $\Voc$ and $\jgen$ are obtained from light intensity-dependent $JV$ curves. (b) The $J(\Vimp)$ curve is constructed using the $\Voc$ and $\jgen$ values in (a), shifted down by the generation current at specific light intensities. (c) External series resistance, $\Rext$, is determined by fitting $dV/dJ$ at high forward bias. The inset shows transport resistance dominating over $\Rext$ at voltages below $\Voc$. (d) Voltage drop ($\Vtr$) and open-circuit conductivity ($\sigmaoc$) are determined by comparing $J(\Vext)$ and $J(\Vimp)$.} 
    \label{fig:tutorial}
\end{figure*}

This section provides a step-by-step guide for determining transport resistance from experimental data. The required data are current--voltage measurements at varying light intensities, as shown in Figure~\ref{fig:tutorial}(a). This easily accessible characterisation technique is often used to determine the light intensity-dependent $\Voc$ and $\jsc$. This same dataset can be employed to evaluate transport resistance, offering a simple and reliable method for distinguishing between fill factor losses caused by transport resistance and those due to recombination. 
This method also allows to determine the figure of merit $\alpha$ and the effective conductivity $\sigma$. 

To analyse transport resistance losses, the illuminated current--voltage curve of a solar cell, $J(\Vext)$, is compared to a $J(\Vimp)$ curve without transport resistance. The latter represents a hypothetical device with perfect charge extraction, and needs to be constructed using the difference of recombination and generation current density: $J(\Vimp) = \jrec(\Vimp) - \jgen$. The recombination current can be determined from the suns-$\Voc$ curve, essentially $\jrec(\Vimp)$, that is typically used to determine the diode ideality factor.\cite{tvingstedt_temperature_2016} A very powerful, alternative approach to evaluate transport resistance losses is by suns-PL measurements. This method works on both devices and even thin films without electrodes.\cite{rau2020luminescence,list_determination_2023} A similar characterisation approach for devices is based on electroluminescence.\cite{faisst2022direct}

In Section~\ref{sec:theory}, we discussed that in a solar cell without transport resistance loss, where $\Vext = \Vimp$, the QFLs remain flat. For a real solar cell this scenario corresponds to the situation under open-circuit conditions in Figure~\ref{fig:theory}. Thus, we can estimate $\Vimp$ using $\Voc$ at various light intensities. Under open-circuit conditions, the net current is zero, and all generated charge carriers recombine. This allows to determine the recombination current density $\jrec$ from the generation current density at different illumination intensities, $\jgen(\Phi)$, as shown in Figure~\ref{fig:tutorial}(a). The latter is approximated using the current density at negative voltage, typically at -0.5~V. Under these conditions, $J\approx\jgen$, as recombination losses are significantly suppressed, hence most of the photogenerated charge carriers are extracted. Increasing the reverse bias improves the accuracy of this approximation. For both, the approximation of $\jgen$ and the corresponding $\Voc$, the parasitic effect of the parallel resistance has to be accounted for: when it is too low, leakage currents through the solar cell limit the reverse bias at which generation current can be determined accurately. In addition, also the light intensity dependent $\Voc$ values are affected by these leakage currents across the parallel resistance at low illumination levels. It is important that only the data which is not influenced by such leakage current is included in the analysis. The appropriate range can be determined by comparing the suns-$\Voc$ data to the dark $JV$ curve. 

The $\jrec$ and $\Vimp$ pairs, which were determined from the illuminated $JV$ curves as described above, are then used to construct the $\jrec(\Vimp)$ curve as shown in Figure~\ref{fig:tutorial}(b). 
The net current density $J$ is calculated as the difference between the recombination and generation current densities: $J(\Vimp) = \jrec(\Vimp) - \jgen$. This means that the recombination current curve is effectively shifted down by $\jgen$ at a specific light intensity. For example, to obtain the $J(\Vimp)$ curve at 1~sun, we shift $\jrec(\Vimp)$ down by $\jgen$ that was determined from the data at 1~sun. This results in the $J(\Vimp)$ curve that represents the transport resistance-free version of the measured $J(\Vext)$ curve of a solar cell at 1~sun. Both curves are shown in Figure~\ref{fig:tutorial}(d), and the voltage difference at a given current density is determined by the total series resistance.

The total series resistance has two contributions: the voltage-dependent transport resistance and the external series resistance of the circuit, $\Rext$. The impact of the latter needs to be accounted for to determine the transport resistance correctly. For that, we choose the conditions under which $\Rext$ is dominant -- at large positive voltage, where transport resistance becomes negligible due to large charge carrier densities and thus large conductivities. Under these conditions, $\Rext$ is estimated using the relation $dV/dJ \approx \nid\kT/e(J+\jgen) + \Rext$, as shown in Figure~\ref{fig:tutorial}(c) for a PM6:Y6 solar cell. The inset of the figure shows a comparison between $\Rext$ and the transport resistance as a function of external voltage. The latter dominates across all voltages in the fourth quadrant of the $JV$ curve, and has a much larger impact on the fill factor compared to $\Rext$. 

This evaluation requires the recombination ideality factor $\nid$, which we determine from the light intensity dependent $\Voc$ as\cite{tvingstedt_temperature_2016}
\begin{equation}\label{eq:nid}
    \nid = \frac{e}{\kT} \bl \frac{\der \ln \Phi}{\der \Voc} \br^{-1} . 
\end{equation}
Here, $\Phi$ is the light intensity, which can be measured using a photodiode or estimated with the calibrated value of the neutral density filters. Alternatively, $\jgen$ can be used instead of $\Phi$ in this expression. It is important to ensure that the data is limited to the region where leakage currents do not interfere, as these typically affect $\Voc$ at low illumination levels. 
 
The voltage drop caused by the transport resistance is determined by comparing the voltages of the two $JV$ curves at the same current density and subtracting the effect of the external series resistance:\cite{schiefer_determination_2014} 
\begin{equation}\label{eq:Vtr1}
    \Vtr(J) = \Vext(J) - \Vimp(J) - J\Rext . 
\end{equation}
To illustrate this method, we compare the measured current--voltage curve of the real solar cell, $J(\Vext)$, to the constructed one of a transport-resistance free device, $J(\Vimp)$, in Figure~\ref{fig:tutorial}(d). When we calculate $\Vtr$, the external and implied voltages are compared at the same $J$ value, which cannot be done using the raw data. Hence, the voltages $\Vext$ and $\Vimp$ have to be interpolated to a new current density axis. This process is more reliable when a high density of data points from the suns-$\Voc$ measurements is used. This can, e.g., be done by using a double neutral density filter wheel in combination with a stable illumination source such as a cw laser diode. For interpolation, the new current density axis is created as a logarithmically spaced set of $J$ values, with its minimum and maximum matching the range of $\jrec$. To align with this new $J$ axis, the experimental $JV$ curves are shifted upward by respective $\jgen$, ensuring all values remain positive, as required for taking the logarithm. Once this transformation is complete, the voltages $\Vext$ and $\Vimp$ can be interpolated onto the logarithm of the new $J$ axis. After the interpolation, the new $J$ axis is shifted down by $\jgen$ for each $JV$ curve separately, to replace the experimentally determined $JV$ curves.

The obtained $\Vtr$ values can be used to extract various parameters. For instance, the effective conductivity at any point on the $JV$ curve can be determined using \cref{eq:Rtr}. 
Except at open-circuit, where this equation no longer applies since the current is zero, the conductivity is directly calculated from the difference in slopes of $\Vext(J)$ and $\Vimp(J)$ following \cref{eq:JV-slope-sigma}. Combining the conductivity at $\Voc$ with the estimated $\jgen$ allows to evaluate the figure of merit $\alpha$ for each light intensity. 
The apparent ideality factor, which appears in the diode equation, is equal to $\nid+\alpha$ at $\Voc$. It can also be derived directly from the slope of the $\Vext(J)$ following \cref{eq:JV-slope-voc}. At any point other than $\Voc$, e.g., at the maximum power point, $\napp$ is calculated using \cref{eq:napp}. 

Finally, fill factor losses can be attributed to two main factors: recombination and transport resistance. Distinguishing these requires both the measured fill factor from the $J(\Vext)$ curve and the pseudo-fill factor $\pFF$, representing the fill factor of a solar cell without transport resistance. The pseudo-fill factor can be obtained from the $J(\Vimp)$ curve or alternatively, approximated using the Green equation, (\cref{eqn:green}). To determine the upper limit of the fill factor, $\FFmax$, we can calculate a $JV$ curve using a diode equation with the same parameters as for $J(\Vimp)$, but setting the ideality factor to unity. Alternatively, the Green equation can be applied with $\nid=1$. The fill factor losses due to recombination are calculated as $\FFmax-\pFF$, while the additional losses caused by transport resistance are equal to $\pFF-\FF$.

\section{What can we learn from the transport resistance figures of merit}\label{sec:FoM}

An often used approach to relate material-specific properties to photovoltaic device performance is to define Figures of Merit (FoM). The oldest and most commonly used FoM to quantify the competition between charge extraction and recombination is the $\mu\tau$ product. It was first introduced by Hecht\cite{hecht_zum_1932} in 1932 to describe the current--voltage characteristics of photogenerated charge carriers in crystalline devices.\cite{crandall1983modeling} Here, $\tau$ is the lifetime of photogenerated carriers. For the classical doped semiconductor crystals, this would be the minority carrier lifetime, which is -- in first order approximation -- independent of illumination intensity and electric field. This FoM was related to the \emph{Schubweg} $\mu\tau F$, a \enquote{drift length} with the electric field $F$. A high value exceeding the crystal thickness corresponded to low recombination losses and a high fill factor. However, this FoM did not consider transport losses, just recombination losses during charge extraction. The Hecht equation has also been empirically applied to organic solar cells.\cite{street_interface_2010} 
However, for nongeminate recombination typical for OSCs, $\tau$ becomes a function of carrier concentration, which renders the $\mu\tau$ product inappropriate to assess the performance of most organic solar cells -- unless $\mu\tau(n)$ and, indeed, transport losses are considered.\cite{deibel_comment_2010,street_reply_2010} 

In year 2016, Neher et al.\ proposed a FoM $\alpha$ for transport-limited photocurrents in organic solar cells, defined as:
\begin{equation}\label{eq:N2}
    \alpha = \sqrt{\frac{e^2 k G d^4}{4\mueff^2\left( \kT\right)^2}} . 
\end{equation}
Here, $k$ is the coefficient under the assumption of second-order recombination, $G$ the generation rate, and $\mueff$ is an effective mobility defined as the geometric mean of the electron and hole mobilities. This expression holds assuming the QFL splitting is constant throughout the entire active layer, meaning that the recombination rate is independent of the position within the active layer. 
Note that \cref{eq:N2} is a special case of \cref{eq:alpha-general} that we defined as the general case in the theoretical section. To arrive at \cref{eq:N2}, the generation current is approximated as $\jgen=eLG$, and at open-circuit, where $\alpha$ is defined, it is equal to the second-order recombination current $\jrec=eLkn^2$. This allows to relate the charge carrier density $n$ to the recombination prefactor $k$, i.e.\ $n=\sqrt{G/k}$. The factor 4 in the denominator comes from assuming equal electron and hole conductivities, which allows to express the effective conductivity as $\sigmaeff=2e\mueff n$.\cite{wurfel_impact_2015} 
Substituting the above into \cref{eq:alpha-general} results in Equation~\eqref{eq:N2}. 
The FoM $\alpha$ represents the competition between recombination and charge transport mechanisms. In organic solar cells, both processes are strong functions of energetic disorder, which is often assumed to be of a Gaussian shape. For this case, the recombination prefactor $k$ and the effective mobility $\mueff$ are independent of light intensity. However, \cref{eq:N2} is also valid for the general case, where energetic disorder is not Gaussian, albeit $k$ and $\mueff$ become \emph{effective} parameters in the sense that they do not only depend on the electron and hole contributions, but also on the respective charge carrier concentrations.

\begin{figure*}[!tb]
    \centering
    \includegraphics[width=1.0\linewidth]{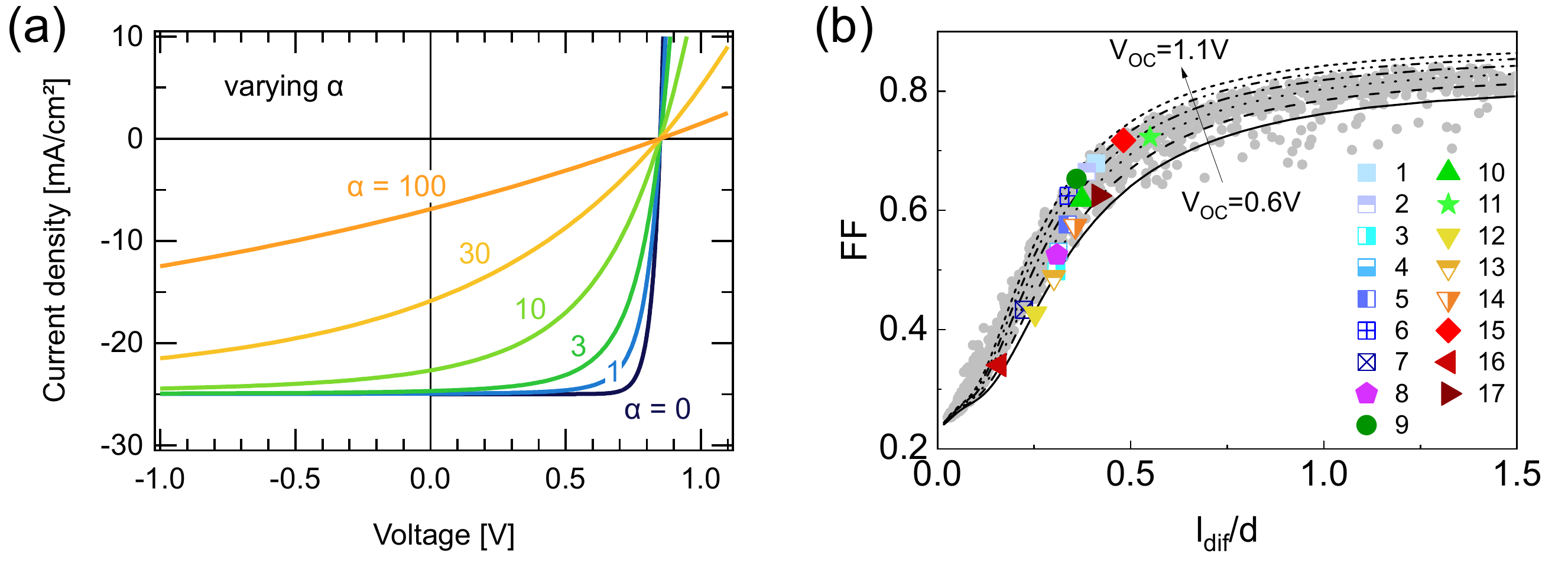}
    \caption{(a) Current--voltage curves calculated using \cref{eq:N6}. The generation current is kept constant, while $\alpha$ is varied from 0 to 100. The FoM $\alpha$ effectively stretches the $JV$ curve relative to $\Voc$. (b) $\FF$ with the relative diffusion length $l_{dif}/d$ at 1~sun illumination for a number of photovoltaic systems. Grey points in the background depict the results of the drift--diffusion simulations for nearly balanced carrier mobilities ($\Voc = 0.7 - 0.9$~V; $\mu_{n}/\mu_{p} = 0.5 - 2.0$.\cite{tokmoldin_explaining_2021}}
    \label{fig:alpha}
\end{figure*}

We note that $\alpha$ is very similar to the FoM $\theta$, introduced by Bartesaghi et al.\ in 2015 to describe the competition between second order nongeminate recombination and charge extraction under short-circuit conditions.\cite{bartesaghi2015competition} The major difference is that $\theta$ uses an internal bias $V_i$ instead of $\alpha$'s $\kT$. $V_i$ is not the implied voltage, but approximated as the difference between the built-in potential given by the electrode work function difference, minus 0.4 V. Later, we show that $\alpha$ can be expressed with respect to the charge carrier diffusion length, whereas $\theta$ relates to drift. To demonstrate the usefulness of the FoM $\theta$, fill factors were extracted from simulated $JV$ curves, with the physical parameters entering $\theta$ being varied over a wide range at room temperature, and shown to display a unique dependence on $\theta$, with little scatter. However, in contrast to $\alpha$, an analytical relation of $\theta$ to device properties was not established. We also note that a potential space-charge limitation of the photocurrent\cite{mihailetchi_space-charge_2005,stolterfoht2015photocarrier} can have an impact on the competition between recombination and extraction,\cite{stolterfoht2016role} but was also not considered in this FoM. Later, Kaienburg et al.\ showed that the device $\FF$ can be related to a so-called collection coefficient $\gamma$,\cite{kaienburg2016extracting} for nearly balanced mobilities. The main difference to the FoMs $\alpha$ and $\theta$ is that while $\alpha$ and $\theta$ are zero for the ideal case without collection losses, $\gamma$ goes towards infinity. 

All of these FoMs have in common that they are roughly, directly or inversely, proportional to $(k G)^{1/2} d^2/\mueff$. This is a direct consequence of the dynamic equilibrium between charge generation, second order recombination (in this special case), and charge extraction. The interplay of these processes under steady state conditions establishes a carrier density $n$. If we consider the current to be a drift current at short-circuit conditions, the result corresponds to the FoM $\theta$. Assuming a diffusion current with the voltage $\kT/e$ instead, we arrive at the FoM $\alpha$. 

The Hecht equation\cite{hecht_zum_1932} from the beginning of this section used the FoM $\mu\tau(n)$ to describe the recombination--extraction interplay for a device with perfect charge extraction. The FoM $\alpha$ that describes, additionally, transport losses is proportional to $\jgen/\sigmaeff$. As it is defined at open-circuit conditions, $\jgen \propto G = R = n/\tau(n)$ where $\tau(n) = 1/(kn)$ for second order recombination, and $\sigmaeff \propto \mueff$. This yields $\alpha \propto (\mueff\tau(n))^{-1}$. Therefore, interestingly, both recombination-to-extraction \emph{and} the additional transport losses are essentially related to the (carrier concentration-dependent) $\mu\tau$-product. 

Next we consider how the FoM $\alpha$ enters the diode equation. Neher et al.\cite{neher2016new} could approximate the implied voltage by an equation containing the externally applied voltage, the open circuit voltage, and an apparent ideality factor $1+\alpha$. The found relation corresponds to \cref{eq:napp} evaluated at $\Voc$ assuming $\nid=1$. This important result led to a modified diode equation for transport-limited photocurrents: 
\begin{equation}\label{eq:N6}
    J(\Vext) = \jgen\left( \exp\left(\frac{e\left(\Vext - \Voc\right)}{(1 + \alpha)\kT}\right) - 1 \right) . 
\end{equation}
This corresponds to \cref{eq:j_Vext_napp} with $\napp = 1 + \alpha$, representing the ideality factor of a transport-limited device around the open-circuit voltage -- as long as the recombination ideality factor can be approximated by a value of $1$.

With \cref{eq:N6}, a way to calculate complete $JV$ curves under illumination was established, with the FoM $\alpha$ determining the shape of the curve.\cite{albrecht2014quantifying} This was demonstrated for a series of blends, where the chemical structure, and with it the hole mobility, was varied.\cite{shoaee2018role}
When $\alpha \ll 1$, $\Vimp$ becomes equal to $\Vext$, $\Vtr$ becomes zero, and \cref{eq:j_Vext_napp} simplifies to the ideal diode \cref{eq:opv-diode} without transport losses. If, however, $\alpha\gg$1, the device current becomes severely transport limited and $\Vimp$ changes much more weakly than $\Vext$, thus causing significant voltage losses $\Vtr = \Vext - \Voc$ even well below $\Voc$. How the FoM $\alpha$ impacts the $JV$ curve of a solar cell is presented in Figure~\ref{fig:alpha}(a), showing that a high transport resistance essentially stretches the illuminated $JV$ curve relative to $\Voc$. For a real device with transport resistance losses, the case with perfect charge extraction, corresponding to $\alpha=0$ in the figure, can be reconstructed by suns-$\Voc$ measurements as described in section~\ref{sec:exp}: at $\Voc$, the current density $J=0$ by definition, and then $\Vimp=\Voc$.

\begin{figure*}[t]
    \centering
    \includegraphics[width=1\linewidth]{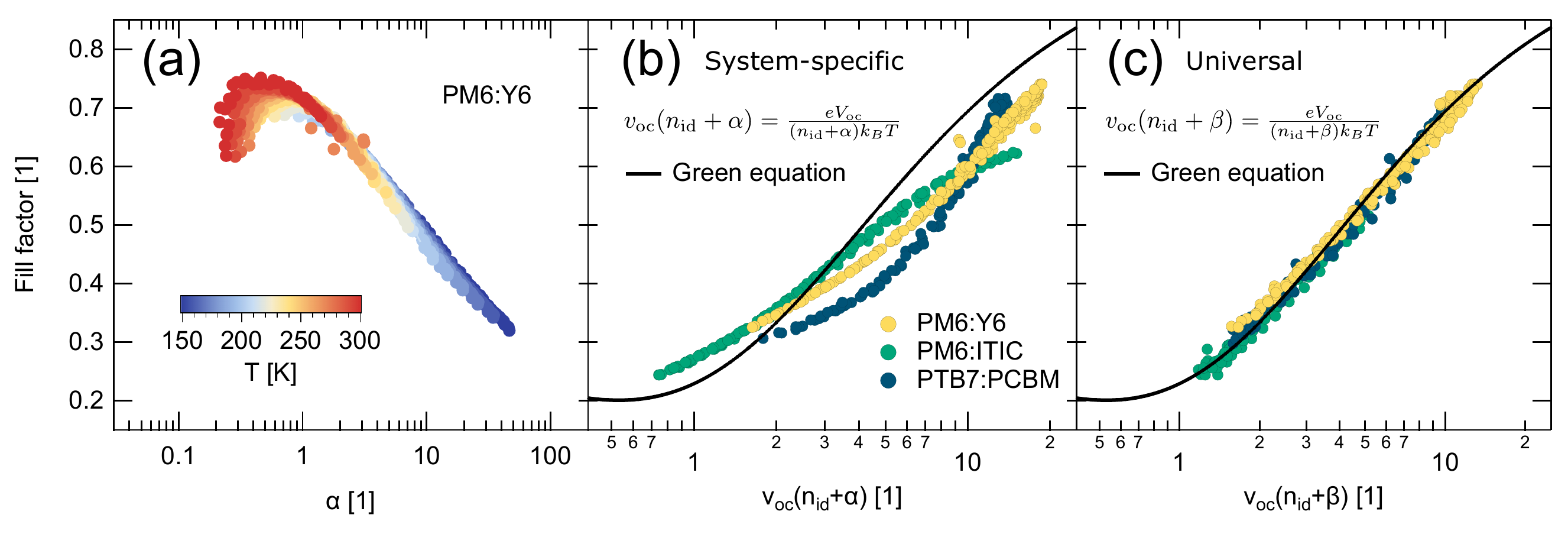}
    \caption{(a) The fill factor of PM6:Y6 shows a good correlation to the FoM $\alpha$ within the 150--300~K temperature range. (b) This correlation is system-specific, also when shown as a function of $\voc$ in comparison to \cref{eqn:green}. (c) However, when $\voc$ is evaluated with an ideality factor $\nid+\beta$, given by \cref{eq:voc}, the fill factor can be universally approximated by the Green equation. The data was taken from Ref.~\parencite{saladina_transport_2024}.}
    \label{fig:alpha-FF}
\end{figure*}

The framework of Neher et al.\cite{neher2016new} also allowed to establish a quantitative relation between $\FF$ and $\alpha$, as shown in Figure~\ref{fig:alpha-FF}(a). For $\alpha \ll 1$, $\FF$ varies between 80\% and 90\%, as predicted by the ideal diode equation. There is a sharp drop of $\FF$ once the transport-limited regime is entered for $\alpha \gg 1$. 
Related to this, \cref{eq:N2} can be used to assess the quality of the active material, namely whether for a given generation current the photocurrent will suffer from strong transport losses or not. For state-of-the-art organic solar cells, $\jgen$ can easily reach 25 mA/cm$^2$ with $d=100$ nm. Then, for a reasonable $k$ of $10^{-11}\,\text{cm}^3/\text{s}$, $\mueff$ must be larger than $10^{-3}\,\text{cm}^2/\text{V}\,\text{s}$ to have $\alpha<1$, which is quite difficult to achieve in organic donor--acceptor blends due to their structural and energetic disorder. 
We point out that $\alpha$ is very sensitive to changes in the effective conductivity or recombination current: a ten percent change in either of the two will change $\alpha$ by roughly ten percent, too. That will already change the fill factor, as shown in Figure~\ref{fig:alpha-FF}(a) for PM6:Y6 solar cell measured at various temperatures. This is particularly important to understand for the optimisation of already very good solar cells -- as discussed in section~\ref{sec:meta-review} -- in striving towards the maximum $\FF$. 
Following the concept introduced in Ref.~\parencite{green1982accuracy} that was already outlined in section~\ref{sec:meta-review}, the $\FF$ can also be represented by a function of the normalised open-circuit voltage, \cref{eqn:green}, with $\napp = 1 + \alpha$. This is shown in Figure~\ref{fig:alpha-FF}(b), parametric in $\Voc$. Also shown are experimental data from different material combinations, to show that $\alpha$ is system-specific and not universally described by the Green equation. Further below, we describe a modified figure of merit, $\beta$, that provides a universal description of organic solar cells with \cref{eqn:green}.

With the assumptions made in \cref{eq:N2}, $\alpha$ can be related to the ratio between the active layer thickness $d$ and the diffusion length $l_\mathrm{dif}$. Assuming the Einstein relation $D/\mu =\kT/e$, this yields\cite{tokmoldin_explaining_2021}
\begin{equation}\begin{split}\label{eq:N10a}
    \alpha 
    &= \frac{d^2}{2D\tau} 
    = \left(\frac{d}{\sqrt{2}l_\mathrm{dif}}\right)^2 . 
\end{split}\end{equation}
Due to the direct relationship between the FoM $\alpha$ and $l_\mathrm{dif}/d$, the ratio of the effective diffusion length and the active layer thickness can be used on its own for correlating with $\FF$. This is shown in Figure~\ref{fig:alpha}(b) for a range of organic solar cells with thicknesses ranging from 80~nm to 550~nm at 1~sun illumination at room temperature. A good overlap of the experimental data both with the analytical expression (\cref{eq:opv-diode-extended}) and the simulation results is observed. This shows that the $\FF$ of most studied blends is limited by insufficient carrier diffusion to the contacts at low internal fields, near $\Voc$. In analogy, the FoM $\theta$ can be expressed as the ratio between the active layer thickness $d$ and the drift length at short-circuit conditions.\cite{tokmoldin_explaining_2021}

The FoM $\alpha$ is inversely proportional to the effective conductivity, as seen from \cref{eq:alpha-general}, and as it contains carrier concentration it is exponentially dependent on the voltage. We outlined already in the introduction that the conductivity decreases when going to voltages smaller than $\Voc$. This makes transport losses at the maximum power point -- and, indeed, the short circuit -- more pronounced. The exponential voltage dependence of the conductivity can be described by an ideality factor $\nsig$, in direct analogy to the recombination ideality factor $\nid$ that represents the exponential dependence of the current density on the voltage (see Equations~\eqref{eq:jgen} and \eqref{eq:sigma} in section~\ref{sec:theory}). Both depend on the mode of recombination and transport, respectively, and are determined by the detailed shape of the density of states in the disordered semiconductors used in OPV.\cite{saladina_power-law_2023, saladina_transport_2024} The original derivation of $\alpha$ by Neher et al.\cite{neher2016new} described above considers only $\nid=1$, which works well to predict the $\FF$ for a given material system at room temperature, as shown in Figure~\ref{fig:alpha-FF}(a). It even works, to some degree, for temperature dependent data, although the $\FF$ is better represented with respect to $\alpha \kT$. Below we will discuss how the FoM can be corrected to be even more accurate. 

An interesting aspect to consider is how the electron and hole mobility (or conductivity) contribute to the effective mobility $\mueff$ (or the effective conductivity $\sigmaeff$) in the transport resistance or the FoM $\alpha$. Since $\sigmaeff$ influences the slope of the $JV$ curve, it is important to understand which charge carrier dominates the effective conductivity. The voltage drop due to transport resistance can be expressed via the individual electron and hole conductivities as:\cite{schiefer_determination_2014}
\begin{equation}\label{eq:Vtr}
	\frac{\Vtr}{d} = \frac{J_p}{\sigma_p} + \frac{J_n}{\sigma_n} = \frac{J}{\sigmaeff}, 
\end{equation}
with the electron and hole currents $J_n = J_p = J/2$, the electron and hole conductivities $\sigma_n$ and $\sigma_p$ in the acceptor and donor phases, respectively, and the effective conductivity given by the harmonic mean of its constituents, i.e., $\sigmaeff = 2/(\sigma_n^{-1}+\sigma_p^{-1})$. \cite{muller_analysis_2013,schiefer_determination_2014,albrecht_quantifying_2014}
Lower conductivity leads to a stronger QFL gradient and a higher contribution to the voltage loss. Consequently, the effective conductivity determined via \cref{eq:JV-slope-sigma} is dominated by the slower charge carrier. Later publications\cite{wurfel_impact_2015,neher2016new} considered the voltage drop to be given by the \emph{geometric} mean of charge carrier \emph{mobilities}, $\mueff = \sqrt{\mu_n \mu_p}$, by assuming identical gradients for the quasi-Fermi levels, which lead to the condition $\sigma_n = \sigma_p$. How electron and hole conductivity really contribute to the effective conductivity still has to be addressed by suitable experiments. It will be interesting to learn if the effective conductivity leading to transport losses will come from a mean of mobilities that is similar to nongeminate recombination, which has been predicted to depend on the spatial extent of the donor--acceptor domains.\cite{heiber_encounter-limited_2015}

The light intensity and temperature dependence of $\alpha$ is implicitly determined by energetic disorder, given its relation to recombination and charge transport. When the effective energetic disorder for electrons and holes is equal, $\alpha$ scales simply with the charge carrier density. 
However, this balance can shift under certain conditions, such as lowering the light intensity, if the distribution of states in energy differs for donor and acceptor. 
When the disorder is no longer equal, the ease of hopping between sites may become different for electrons and holes. For instance, a recent study revealed that the energetic distribution of states in modern organic solar cells can be described as a combination of a Gaussian and an approximately exponential distribution found to correspond to a power-law density of states.\cite{saladina_power-law_2023} In systems with such mixed density of states, reducing the QFL splitting exposes charge carriers to greater energetic disorder, and more importantly, this disorder is in general not equal for electrons and holes. Consequently, the contributions of electrons and holes to recombination and charge transport are also different. How these processes depend on the charge carrier concentration -- and therefore on the density of states -- is accounted for in the general equations for $\jgen$ and $\sigma$ (Equations~\eqref{eq:jgen} and \eqref{eq:sigma}) through their respective ideality factors. The ratio of the ideality factors $\nid/\nsig$ determines the exponential voltage dependence of $\alpha \propto \jgen/\sigmaeff$, offering insights into which density of states dominates recombination and transport losses. For example, for a mixed density of states, the dependence of $\alpha$ on light intensity with the power of $1.5 - \nid$ implies that transport is limited by charge carriers from the power-law DOS, while the major recombination channel is governed by carriers from the Gaussian DOS.\cite{saladina_transport_2024} Hence, considering ideality factors in $\alpha$ leads to a deeper understanding of the physical mechanisms within a solar cell, as it satisfies a more general case that is valid not only for equal energetic disorder, but also for cases when contribution of electrons and holes to recombination and transport are not perfectly balanced. For example, this general relation can be employed for the temperature-dependent studies or the studies of material properties and active layer morphology. 

While the FoM $\alpha$ is able to predict the fill factor well, the apparent ideality factor -- approximated by $1+\alpha$ -- is only valid around $\Voc$. At the maximum power point, where the $\FF$ is defined, $\alpha$ underestimates the transport resistance -- and therefore the $\FF$ loss. Saladina and Deibel\cite{saladina_transport_2024} presented an approach to predict an improved ideality factor $\beta$, which allows to predict the transport resistance with the knowledge of $\alpha$, $\nid$ and $\nsig$. To do so, they proposed an iterative approach. The resulting $\beta$ (called so because it works better) can be used also in the normalised open-circuit voltage $\voc$. Thus, replacing the apparent ideality factor $\napp$, with the sum of $\nid$ and the iteratively determined $\beta$ (\cref{eq:napp}), we state
\begin{equation}\begin{split}\label{eq:voc}
    \voc &= \frac{e\Voc}{(\nid + \beta)\kT} , \quad\text{where} \\
    \beta &= \alpha \cdot \frac{\voc \bl \voc + 1 \br^{\frac{\nid}{\nsig} - 1}}{\ln\bl \voc + 1 \br} . 
\end{split}\end{equation}
With this modified definition, the real device $\FF$ can be predicted with the original empirical equation by Martin Green,\cite{green1982accuracy} \cref{eqn:green}, with high accuracy for different material systems across a wide range of temperatures and light intensities (Figure~\ref{fig:alpha-FF}(c)).\cite{saladina_transport_2024} In contrast, according to Figure~\ref{fig:alpha-FF}(b), $\alpha$ can only predict the fill factor well for a single given system. The FoM $\voc$, as defined in \cref{eq:voc}, can furthermore be used to link the collection efficiency at the maximum power point to the transport resistance in a very simple manner,\cite{saladina_transport_2024} yielding
\begin{equation}
    \etacol = \frac{J}{\jgen} = \frac{\voc}{\voc + 1} .
\end{equation}
The transport resistance also predicts the current losses around short-circuit conditions that are called \enquote{photoshunt} losses in the literature. The photoshunt is described in more detail in section~\ref{sec:Rphoto}.

\section{Different perspectives on the transport resistance}\label{sec:perspectives}

Relating the photocurrent and photovoltage for a given irradiation is one of the central tasks of solar cell device physics. Most physical models start with a generation--recombination framework, where the current at a given voltage and irradiance is split up into photogeneration and recombination terms. A simple version of such a model can be expressed as $J={J}_\mathrm{rec}(V, \Phi)-{J}_\mathrm{gen}(\Phi)$, where ${J}_\mathrm{rec}(V, \Phi)$ is the recombination current density that depends on voltage $V$ and illumination intensity $\Phi$, while the photogenerated current density ${J}_\mathrm{gen}(\Phi)$ only depends on the absorbed photon flux. Despite its advantages, this description presents some challenges: there is no universal and obvious relation between ${J}_\mathrm{rec}(V, \Phi)$ and $V$, nor an obvious connection of ${J}_\mathrm{rec}(V, \Phi)$ to inefficient charge carrier collection due to slow charge transport through the semiconducting layers of a solar cell. 
Both problems are interrelated, and various approaches in the field of photovoltaics address charge collection losses. These approaches differ in how they account for collection losses and the parameters used to create analytical variants of the current--voltage equation linking ${J}_\mathrm{rec}(V,\Phi)$ and $V$. To maintain the familiar one-diode model structure with parasitic parallel and series resistances, one could (i) incorporate transport resistance into series resistance (section~\ref{sec:theory-vtr}), (ii) adjust or measure the apparent parallel resistance under illumination (sections~\ref{sec:theory-slopes}, \ref{sec:Rphoto}), (iii) modify the ideality factor (sections~\ref{sec:theory-vtr}, \ref{sec:FoM}), or (iv) introduce a voltage-dependent current source into the circuit (see e.g.\ Ref.~\parencite{crandall1983modeling}).

\subsection{The apparent \enquote{photoshunt}}\label{sec:Rphoto}

A peculiar aspect of transport-related losses is that they affect the illuminated $JV$ curve in ways that are similar to increased series but also reduced parallel resistances. The reason for this effect is that low mobilities in absorber or transport layers lead to significant splitting of the QFLs even down to short circuit. This QFL splitting corresponds to electron and hole concentrations that build-up in the absorber. As recombination rates generally scale with the electron or hole densities, the build-up of charge carriers within the absorber leads to additional recombination currents under illumination that would not be present at the same externally applied voltage in the dark. These recombination currents often have a weak dependence on the externally applied voltage and \emph{appear} as parallel resistance with a linear light intensity dependence. This phenomenon is referred to as \enquote{photoshunt}; however, as we showed in section~\ref{sec:theory-slopes}, it is completely unrelated to the shunt in the dark and only appears under illumination. It has a variety of consequences both for understanding and for quantifying the transport-related recombination losses.

\begin{figure*}[!tb]
    \centering
    \includegraphics[width=0.95\textwidth]{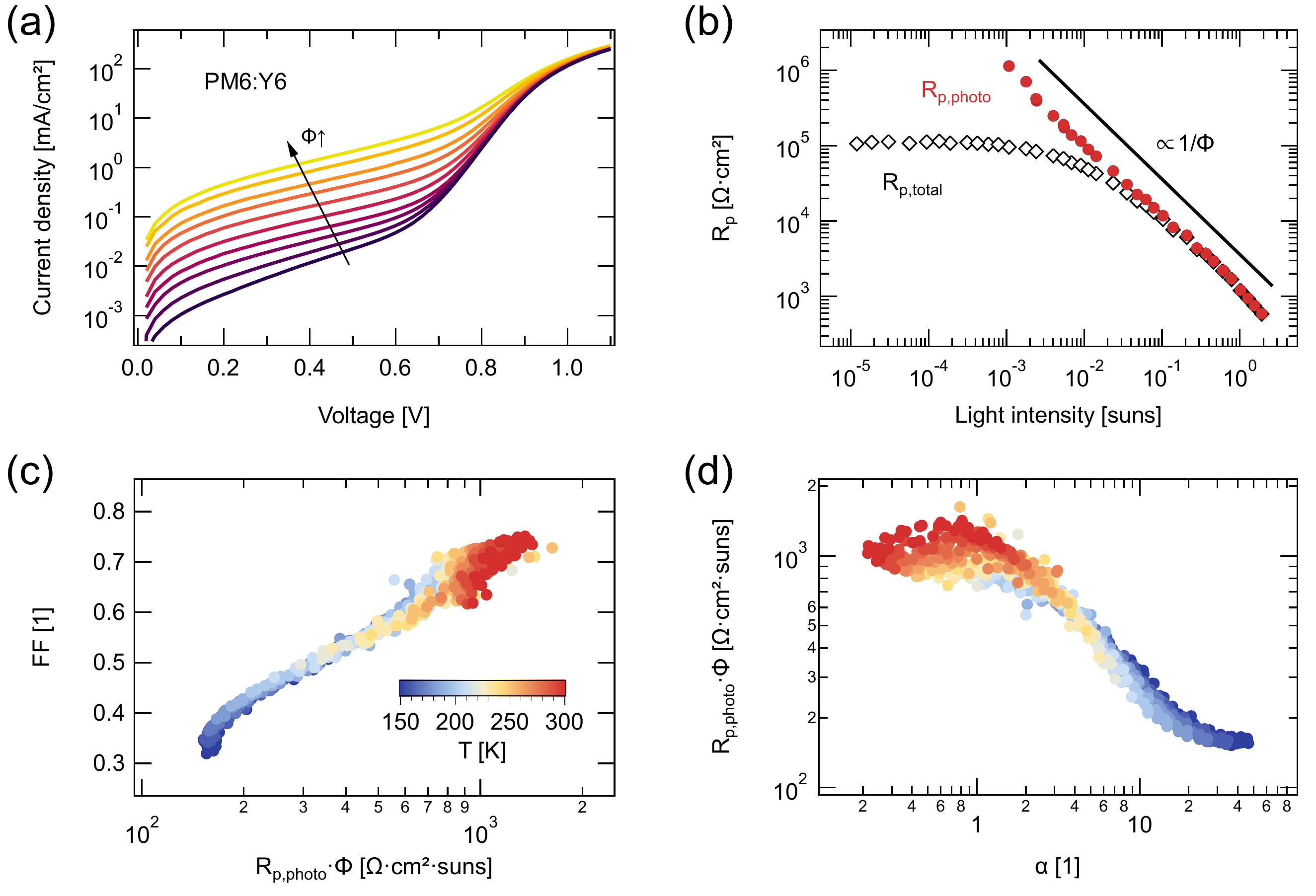}
    \caption{(a) Shifted $JV$ curves of the PM6:Y6 solar cell at 300~K. The near-ohmic current at low forward bias increases significantly with light intensity $\Phi$, indicating an apparent light-intensity-dependent parallel resistance. (b) Photoshunt $\Rphoto$ and total parallel resistance $\Rtotal$ of PM6:Y6, derived from the $JV$ characteristics in (a), using the method from \parencite{kim_correlating_noyear}. At low $\Phi$, $\Rtotal$ saturates to the value of $R_\mathrm{p,dark}$, whereas the corrected $\Rphoto$ does not. The product $\Rphoto \, \Phi$ of PM6:Y6 shows strong correlation to (c) the fill factor, and (d) the FoM $\alpha$ within a temperature range from 150~K to 300~K. The data was taken from Ref.~\parencite{saladina_transport_2024}.} 
    \label{fig:photoshunt}
\end{figure*}

The photoshunt may lead to confusion in interpreting device parameters. Since $JV$ curves are used to determine the resistive elements $\Rs$ and $\Rp$, it is important to understand that $\Rp$ in the dark differs entirely from $\Rp$ at, e.g., 1~sun illumination. To illustrate this effect across different light intensities, we use shifted $JV$ curves plotted on a semi-logarithmic scale, as introduced by Robinson et al.,\cite{robinson1994departures} instead of the traditional linear $JV$ curves under illumination. Each shifted $JV$ curve is constructed by adding $J_\mathrm{sc}$ so they pass through the origin ($V = 0$, $J= 0$). 
Figure~\ref{fig:photoshunt}(a) shows an example for a PM6:Y6 solar cell. The first observation is that light intensity has a systematic effect on the shifted $JV$ curves, even after removing its primary influence -- namely the linear dependence of ${J}_\mathrm{sc}$ on light intensity.\cite{luebke2023understanding} The shifted curves display a diode region (linear on the semi-logarithmic plot) with exponential current--voltage dependence and a lower-voltage part that appears logarithmic on the semi-logarithmic plot, making it approximately linear in reality. This linear relation at small forward bias is the signature of the photoshunt.\cite{grabowski_fill_2022} The curves are offset relative to each other by an additive term which implies (on the logarithmic y-axis) that the current changes by a factor depending on light intensity. 

If we now determine the apparent parallel resistance $\Rtotal$ at every light intensity via 
\begin{equation}\label{equ:Rp}
    \Rtotal=\left.\frac{dV}{dJ}\right|_{V=0},
\end{equation}
we note two contributions \cite{kim_correlating_noyear}: a constant offset originating from the dark parallel resistance and a linearly light-intensity dependent contribution from inefficient charge carrier collection. Figure \ref{fig:photoshunt}(b) illustrates this effect for the PM6:Y6 cell, with the squares showing the total parallel resistance $\Rtotal$, while the circles show the contribution of just the photoshunt. This $\Rphoto$ was determined by
\begin{equation}\label{equ:Rpphoto}
    \Rphoto^{-1} = \Rtotal^{-1}-R_\mathrm{p,dark}^{-1},
\end{equation}
with $R_\mathrm{p,dark}$ corresponding to the dark parallel resistance was estimated at low light intensity of 0.6~msuns, where the total parallel resistance saturates to a constant value. 

It is worth emphasising that $\Rphoto$ is inversely proportional to the light intensity $\Phi$ \cite{kim_correlating_noyear}. To understand why, we must revisit section~\ref{sec:theory}. With the external voltage being zero at short-circuit conditions, the implied voltage becomes the negative of the voltage drop, $\vimp=-\Vtr$. Therefore, we can rewrite \cref{eq:Rtr} for short-circuit conditions as $d/\sigmasc = \Vimp/\jsc$. Next, considering the slope at 0~V, given by \cref{eq:JV-slope-jsc}, which describes the photoshunt, and replacing the conductivity term with the above result, we obtain:
\begin{equation}\label{eq:photoshunt}
    \Rphoto \propto \frac{\Vimp}{\jsc}\cdot\frac{\jgen}{\jgen-\jsc} . 
\end{equation}
Both the generation current and the short-circuit current depend linearly on light intensity, while the implied voltage corresponding to the QFL splitting follows a logarithmic function of $\Phi$. The photoshunt will be dominated by the contribution that changes more rapidly with $\Phi$. Since the logarithmic function grows much slower than the linear one for the same change in input, the photoshunt is dominated by the terms linear in $\Phi$ -- namely $\jgen$ and $\jsc$. 

The photoshunt, which reflects an apparent increase in the slope of the $JV$ curve at 0~V due to transport-related losses, is directly related to the fill factor. 
In section~\ref{sec:theory-slopes}, we noted that the slope alone is insufficient to predict the fill factor; it must be scaled by the generation current (or light intensity). At open-circuit, this scaling transforms the $JV$ slope into the figure-of-merit $\alpha$, which, as demonstrated earlier in Figure~\ref{fig:alpha-FF}(a), predicts the fill factor with remarkable accuracy. Applying similar reasoning to the photoshunt, we multiply it by the light intensity. We then correlate the resulting product, $\Rphoto\Phi$, with the fill factor of the PM6:Y6 solar cell, as shown in Figure~\ref{fig:photoshunt}(c). The fill factor can be described as the logarithm of the product $\Rphoto\Phi$, showing a consistent linear relationship across a wide range of light intensities. Notably, this trend is the same for all the temperatures, indicating that the product $\Rphoto\Phi$ serves as a reliable predictor of the fill factor. Hence, similar to $\alpha$, it can be regarded as a figure-of-merit for fill factor losses. 

Since both $\alpha$ and $\Rphoto\Phi$ can predict the fill factor, they must be related to each other. The slope at any point on the $JV$ curve, described by \cref{eq:JV-slope-general}, can be connected to the FoM $\alpha$ by factoring out the term $\jgen/\sigmaoc$. This leads to a relationship between the photoshunt and $\alpha$, expressed as:
\begin{equation}\label{eq:Rphoto-vs-alpha}
    \Rphoto \cdot \jgen \approx \frac{\alpha\kT}{e}\cdot \frac{\nid}{\nsig}\cdot \bl\frac{\sigmaoc}{\sigmasc}\br^{\frac{\nsig}{\nid} + 1} . 
\end{equation}
Figure~\ref{fig:photoshunt}(d) illustrates this relationship for the PM6:Y6 solar cell. Previously, we showed how the effective conductivities under open-circuit and short-circuit conditions can be evaluated using the corresponding slopes (see section~\ref{sec:theory-slopes}). It is therefore evident that the two FoMs in the above expression are related through the ratio of these conductivities. This relationship is influenced by the ideality factors, with the term $\nsig/\nid$ adjusting the conductivity ratio based on the relative impact of recombination and transport processes. 

To conclude, the photoshunt is an easy to determine parameter that can support any analysis of charge transport and/or $\FF$ losses in various solar cells as long as it is compared at equal light intensity.

\subsection{Photocurrent collection loss instead of voltage loss}\label{sec:currentloss}

\begin{figure*}[!tb]
    \centering
    \includegraphics[width=1\textwidth]{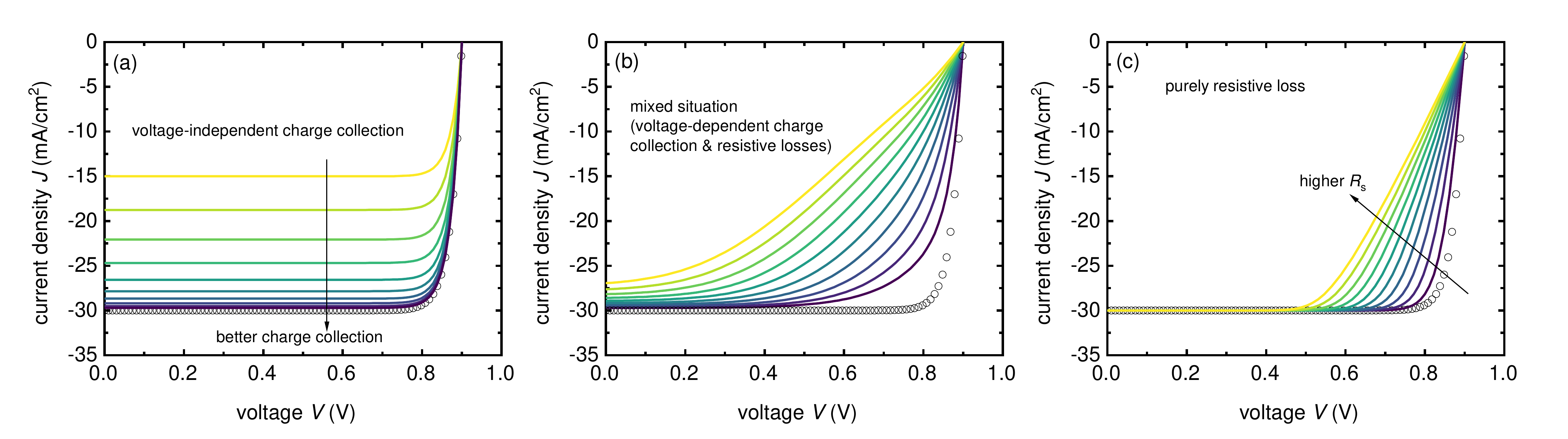}
    \caption{Depending on the electrostatics of the device geometry, charge-transport losses can have different effects on the current--voltage curves of a solar cell. They can either modulate the photocurrent as shown in panel (a), be primarily of resistive nature as illustrated in panel (c), or be a (fairly arbitrary) mixture of the two as shown in panel (b). (a) describes a situation where $\exp\left(\frac{e\Vext}{\kT}\right)-\exp \left(\frac{e\Vimp}{\kT}\right)=\text{const}$ holds, while (c) describes a situation where $\Vext-\Vimp=\text{const}$ holds. In (b) neither of the two terms is constant. Panels (a) and (c) constitute the extreme ends of a spectrum of possible effects of poor charge transport, whereby all real solar cells will occupy a spot somewhere between these two extremes.} 
    \label{fig:Breitenstein01}
\end{figure*}

The concept of transport resistance links the chemical potential of electron--hole pairs (i.e., the QFL splitting $\EF$) within a photovoltaic absorber with the electrostatic potential difference (i.e., the external voltage $\Vext$) between the contacts. 
This is achieved via an additive term that represents the voltage difference caused by transport resistance. Although this is one way to describe the phenomenon, alternative approaches exist, such as the mathematical description developed initially by Breitenstein\cite{breitenstein2014alternative} and later refined by Rau et al.~\cite{rau2020luminescence} This approach uses the difference between $\exp(e\Vext/\kT)$ and $\exp(\EF/\kT)$ instead of the one between $e\Vext$ and $\EF$. It originates from solving the current density equation as a function of the QFL gradient within a space-charge region of a pn-junction. Assuming a parabolic shape of the conduction and valence band edge leads to
\begin{align}
J=J_\mathrm{exc}\left(\exp\left(\frac{e\Vext}{\kT}\right)-\exp \left(\frac{e\Vimp}{\kT}\right)\right) 
\label{equ:Breitenstein01}
\end{align}
whereby the exchange current density $J_\mathrm{exc}$ depends on the ratio of conductivity and Debye length, but is independent of voltage or electric field.\cite{rau2020luminescence} This implies that the finite speed of charge transport through a space-charge region can be expressed with a voltage-independent term that differs in its mathematical description from the transport resistance. The idea of \cref{equ:Breitenstein01} is to capture the physics of charge carrier extraction and injection. If $e\Vext > \EF=e\Vimp$, the injection of charge carriers from the contacts to the absorber dominates. If $e\Vext < \EF=e\Vimp$, charge carrier extraction from the absorber to the contacts dominates. At $e\Vext = \EF=e\Vimp$, we are at open circuit, the Fermi levels are flat, and injection and extraction are perfectly balanced, resulting in zero net current.

We could now equate the generation--recombination model of the current--voltage curve of a solar cell with this extraction--injection type model defined by \cref{equ:Breitenstein01}. This leads to 
\begin{align}\begin{split}
&{J}_\mathrm{0}\left(\exp\left(\frac{{e\Vimp}}{\kT}\right)-1\right)-{J}_\mathrm{gen} \\
&\quad = J_\mathrm{exc}\left(\exp\left(\frac{e\Vext}{\kT}\right)-\exp \left(\frac{e\Vimp}{\kT}\right)\right) 
\label{equ:Breitenstein02}
\end{split}\end{align}
where ${J}_\mathrm{gen}$ is the maximum photogenerated current density, i.e., the ${J}_\mathrm{sc}$ one would have if all charge carriers were collected and none recombined at short circuit. The left hand side of \cref{equ:Breitenstein02} is similar to the one diode model (\cref{eq:opv-diode}) safe for the fact that we used ${J}_\mathrm{gen}$ rather than ${J}_\mathrm{sc}$ and $\Vimp$ rather than $\Vext$. Thereby, the equation does not assume the absence of recombination losses at short circuit but explicitly accounts for their possibility. If we now eliminate $\Vimp$ from \cref{equ:Breitenstein02}, we obtain\cite{rau2020luminescence, kruckemeier23am}
\begin{align}
J=\frac{J_\mathrm{exc}}{J_\mathrm{exc}+J_\mathrm{0}}\left({J}_\mathrm{0}\left(\exp\left(\frac{{e\Vext}}{{k}_\mathrm{B}T}\right)-1\right)-{J}_\mathrm{gen}\right).
\label{equ:Breitenstein03}
\end{align}
Thus, we now have an equation that relates the two observables, current density $J$ and external voltage $\Vext$, to one another. The ratio $\frac{J_\mathrm{exc}}{J_\mathrm{exc}+J_\mathrm{0}}$ plays the role of a collection efficiency that modulates both the recombination current density ${J}_\mathrm{0}\left(\exp\left(\frac{{e\Vext}}{{k}_\mathrm{B}T}\right)-1\right)$ and the photogenerated current density ${J}_\mathrm{gen}$. The intuitive meaning of the fraction $\frac{J_\mathrm{exc}}{J_\mathrm{exc}+J_\mathrm{0}}$ is that as long as charge-carrier exchange between contact and absorber expressed via $J_\mathrm{exc}$ is fast relative to recombination expressed by the term $J_\mathrm{0}$ (that is $J_\mathrm{exc}\gg J_\mathrm{0}$), the collection of charge carriers will also be efficient. In the ideal case observed in a pn-junction, ${J}_\mathrm{exc}$ will be independent of voltage and the loss will be entirely in short-circuit current density while the fill factor would not be affected. In most solar cells that differ from a simple pn-junction, we expect a voltage dependence of ${J}_\mathrm{exc}$ that will lead to a reduction in ${J}_\mathrm{sc}$ and in $\FF$. The exact voltage dependence of ${J}_\mathrm{exc}$ that may well describe organic photovoltaics is so far still a question of active research. 

A significant part of the possible confusion caused by the different models used to describe transport losses is that the different mathematical descriptions may describe the same or different physical realities. Let us make an example: In solar cells with absorber layers that are primarily electric field-free such as silicon, Cu(In,Ga)Se$_2$, and (due to ionic screening) halide perovskites, a voltage-independent or weakly voltage-dependent charge collection loss may result from low conductivities of transport layers or space-charge regions \cite{rau2020luminescence, kruckemeier23am}. The resulting current--voltage curves would behave as shown in Figure~\ref{fig:Breitenstein01}(a), featuring a loss in short-circuit current but not in fill factor. Mathematically, we could write that $\exp\left(\frac{e\Vext}{\kT}\right)-\exp \left(\frac{e\Vimp}{\kT}\right)=\text{const}$. Furthermore, the same type of cells might show resistive losses, e.g. due to very thin metal fingers employed or insufficiently conductive TCOs. In this case, we will observe a purely resistive voltage loss as illustrated in Figure~\ref{fig:Breitenstein01}(c) that mathematically corresponds to $\Vext-\Vimp=\text{const}$. In these two examples different mathematical descriptions are used to describe different physical realities (ohmic losses in the contact layers vs.\ low mobilities within the space charge regions of the semiconductor layers within the actual solar cell). In organic solar cells, however, the physical reality is nearly always a loss caused by the low mobility of a partly or largely depleted active layer. The mathematical description could now be given by voltage dependent transport resistances or voltage dependent collection efficiencies as shown in Figure~\ref{fig:Breitenstein01}(b). There is so far no model that avoids the use of voltage (and light intensity) dependent parameters such that neither of the two cases shown in Figure~\ref{fig:Breitenstein01}(a) and (c) comes close to the reality of organic photovoltaics. However, both of the descriptions can be modified in a way that they can account for the physical reality of low mobility absorber materials. Thus, while there is no unique solution to the question of how to mathematically describe the transport losses, there are alternative solutions that do the job in slightly different ways.

\section{Beyond transport resistance losses}\label{sec:beyond}

In this section, we will consider two loss mechanisms that also influence the fill factor -- electric field-dependent charge photogeneration and recombination of photogenerated with injected charge carriers -- that have a similar appearance as the transport resistance, but have to be considered separately. Both of these processes are (quasi-) first order processes and therefore independent of light intensity.

\subsection{Photogeneration}\label{sec:photogeneration}

The photogeneration process in organic solar cells can depend on the applied voltage and thus impacts the current--voltage characteristics. Since the intermediate state in the photon-to-electron conversion process -- the charge-transfer (CT) state -- dissociates more easily under a stronger electric field,\cite{deibel2010polymer,clarke2010charge} the generation current becomes a function of externally applied voltage. This process was considered until recently one of the primary contributors to the fill factor loss in organic solar cells.\cite{mandoc2007origin,dibb2013limits} However, already in P3HT:PCBM\cite{kniepert_conclusive_2014} and PM6:Y6,\cite{perdigontoro_barrierless_2020} at least at room temperature the photogeneration was virtually field-independent. Nevertheless, for less efficient solar cells or temperature dependent studies, we will explore the impact of field-dependent charge photogeneration on the slope of the $JV$ curve and the fill factor. First, we will define the generation current through the field dependence of CT separation, and then integrate it into the diode equation. 

We consider charge photogeneration proceeding from singlet excitons to separated electrons and holes via the CT state. The probability of dissociation under steady-state conditions can be expressed as\cite{braun1984electric}
\begin{equation}
    \etadiss(F) = \frac{k_d(F)}{k_d(F) + k_f} , 
\end{equation}
with $k_f$ being the recombination to the ground state constant, and $k_d(F)$ is the dissociation rate constant that depends on the electric field $F$. 
When the field is zero, the dissociation probability depends on the CT binding energy, i.e., the energy barrier for separation into free charges.\cite{onsager1938initial,noolandi1982theory} When the field is applied it effectively lowers the barrier and increases the CT dissociation rate constant:\cite{popovic1979electric,giebink2010ideal_ii_role,saladina_charge_2021}
\begin{equation}\begin{split}\label{eq:field-kdF}
    k_d(F) &= k_d(0)\exp\bl \frac{eFr}{\kT} \br . 
\end{split}\end{equation}
Here, $r$ is the effective separation distance of an electron--hole pair. The electric field is approximated as $F\approx (\Vbi-\Vext)/d$, where $\Vbi$ is the built-in potential essentially given by the difference of the electrode work functions. Note that $k_d(F)$ depends on the specific electron--hole separations and their orientations within the electric field. Thus, $k_d(F)$ represents a dissociation constant that is spatially and energetically averaged. With this, the generation current will depend on the generation rate of CT states, and their dissociation probability: 
\begin{equation}\begin{split}\label{eq:field-jgen}
    \jgen 
    &= eL G_\mathrm{CT} \cdot \etadiss(F) , 
\end{split}\end{equation}
with the charge photogeneration rate $G = G_\mathrm{CT} \cdot \etadiss(F)$. Note that the possibility of singlet exciton repopulation will be contained in $G_\mathrm{CT}$, and the nongeminate recombination rate in the prefactor $k_d(0)$. 

Adapting the diode equation to incorporate voltage-dependent charge photogeneration presents two key challenges. First, it is not easy to be implemented analytically. One approach involves using a series expansion of Onsager's result at low electric fields.\cite{onsager1938initial,pai1975onsager} Another one considers only the field-dependent modification of the electrostatic potential, as we have done above, neglecting the impact of the field on electron and hole diffusion.\cite{popovic1979electric,giebink2010ideal_ii_role,saladina_charge_2021} The latter method has been previously employed, for example by Giebink et al., to modify the diode equation for scenarios with voltage-dependent charge photogeneration.\cite{giebink2010ideal_i_derivation} 
The second challenge concerns relating the external and implied voltages to one another. The generation current -- and consequently the charge carrier density and implied voltage -- depends on the electric field that the external voltage creates. This relationship directly affects conductivity and transport resistance, which stretches the $JV$ curve relative to the open-circuit voltage, as was discussed in section~\ref{sec:FoM}. Hence, not only the conductivity influences this stretching, but also the externally applied voltage itself. 

\begin{figure}[!t]
    \centering
    \includegraphics[width=0.95\linewidth]{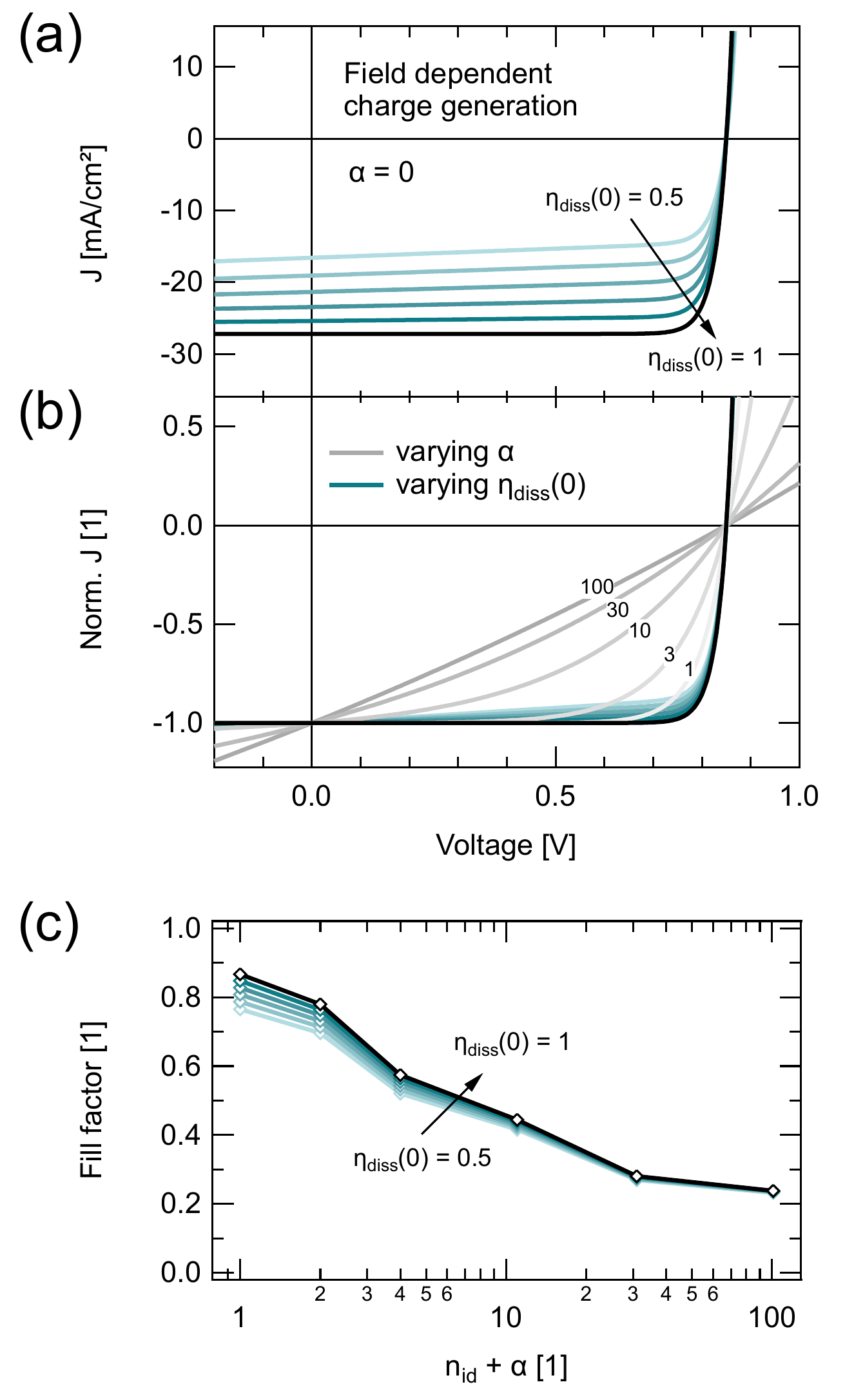}
    \caption{Influence of charge photogeneration on the $JV$ curves, calculated using \cref{eq:field-JV}. (a) Varying $\etadiss(0)$ with $\alpha=0$ to eliminate transport related losses. In (b), $J$ is normalised to $\jsc$ to show the impact of $\etadiss(0)$ and $\alpha$ on the $\FF$. (c) $\FF$ as a function of $\etadiss(0)$ and $\alpha$ with $\nid=1$, showing the dominance of transport losses over those related to charge generation.}
    \label{fig:field}
\end{figure}

To incorporate the voltage-dependent generation current into the diode equation, we rely on approximations. Starting from \cref{eq:j_Vext_napp}, we substitute the above expressions for $\jgen$. Since we are lacking a direct link between $\Vext$ and $\Vimp$, we cannot determine the apparent ideality factor explicitly. Instead, we approximate it as $\nid+\alpha$, assuming it remains constant across the entire $JV$ curve, which leads to: 
\begin{equation}\begin{split}\label{eq:field-JV}
    &J = \jgen(\Voc) \\
    &\cdot \left[ \exp\bl \frac{e\bl\Vext - \Voc\br}{(\nid+\alpha) \kT} \br - \frac{\etadiss(\Vext)}{\etadiss(\Voc)} \right] . 
\end{split}\end{equation}
This approximation underestimates the impact of transport resistance on the fill factor, as we discussed in \cref{sec:FoM}. Consequently, our analysis compares charge generation losses to the lower limit of transport resistance losses. 

We used the above equation to estimate how much of the fill factor losses can be attributed to inefficient charge generation and to assess whether the impact on the fill factor is comparable to the one caused by transport resistance. To begin, we consider a solar cell without transport resistance losses, assuming $\alpha=0$. The parameters for these calculations were selected to be simple, but align with the $\jsc$ and $\Voc$ values typical of state-of-the-art solar cells. We set $\Vbi = 1.16$~V, $\Voc=0.85$~V, $d=100$~nm, $\nid=1$, the CT radius $r$ to 1~nm, and the CT lifetime $(k_d(0)+k_f)^{-1}$ to 2.5~ns.

A decrease in the dissociation efficiency reduces the generation current, as shown in Figure~\ref{fig:field}(a). This shifts the $JV$ curve upwards compared to the ideal case with perfect CT dissociation yield, resulting in an overall loss in performance. Additionally, it alters the slope of the curve around 0~V, but not at open-circuit. While the fill factor is clearly affected, the simultaneous change in $\jsc$ makes it difficult to say by how much. 

To clarify the impact on the fill factor, we normalised the current density to the value of $\jsc$ for each curve in Figure~\ref{fig:field}(b). The influence of the field-dependent charge photogeneration on the shape of the $JV$ curve is apparent, but remains relatively minor. In comparison, increasing transport resistance losses (by raising the figure of merit $\alpha$ from 1 to just 3) produces significantly greater fill factor losses than those theoretically caused by inefficient CT separation. Transport resistance not only affects the $JV$ curve more negatively overall, but also specifically alters the slope near $\Voc$, where $\etadiss$ has virtually no influence. 

The resulting fill factors are shown in Figure~\ref{fig:field}(c). When transport resistance losses are significant ($\alpha>10$), the impact of CT dissociation yield on the fill factor becomes almost negligible. However, a key question is whether high-efficiency state-of-the-art solar cells really experience fill factor losses due to charge photogeneration? For low $\nid+\alpha$ values, ranging from 1 to 2, the voltage dependence of CT dissociation can decrease the fill factor by approximately 10 percent points as $\etadiss(0)$ drops from 1 to 0.5 -- this would be a considerable loss. 

Nonetheless, state-of-the-art solar cells with fill factors as high as 0.7 and EQEs of only 50\% (assuming other losses unrelated to CT dissociation are suppressed) are unheard of, unless they are caused by too thin active layers with respect to the absorption length. Devices that limited typically exhibit low fill factors of 0.5-0.6,\cite{sun2020HighEfficiencyPolymerb,zhou_high-efficiency_2018} indicating pronounced transport and recombination losses. These solar cells are likely to correspond to $\nid+\alpha$ values between 3 and 10 in our example, where CT dissociation is no longer the primary constraint on the fill factor. 

In short, even in solar cells that are limited by a field-dependent photogeneration, the impact of the transport resistance on the fill factor is usually dominant. Since the photogeneration yield is independent of light intensity, whereas transport losses become more pronounced at stronger illumination, the two processes can be mostly distinguished.

\subsection{Recombination of photogenerated charge carriers with injected charge carriers}

\begin{figure*}[!tb]
    \centering
    \includegraphics[width=0.9\textwidth]{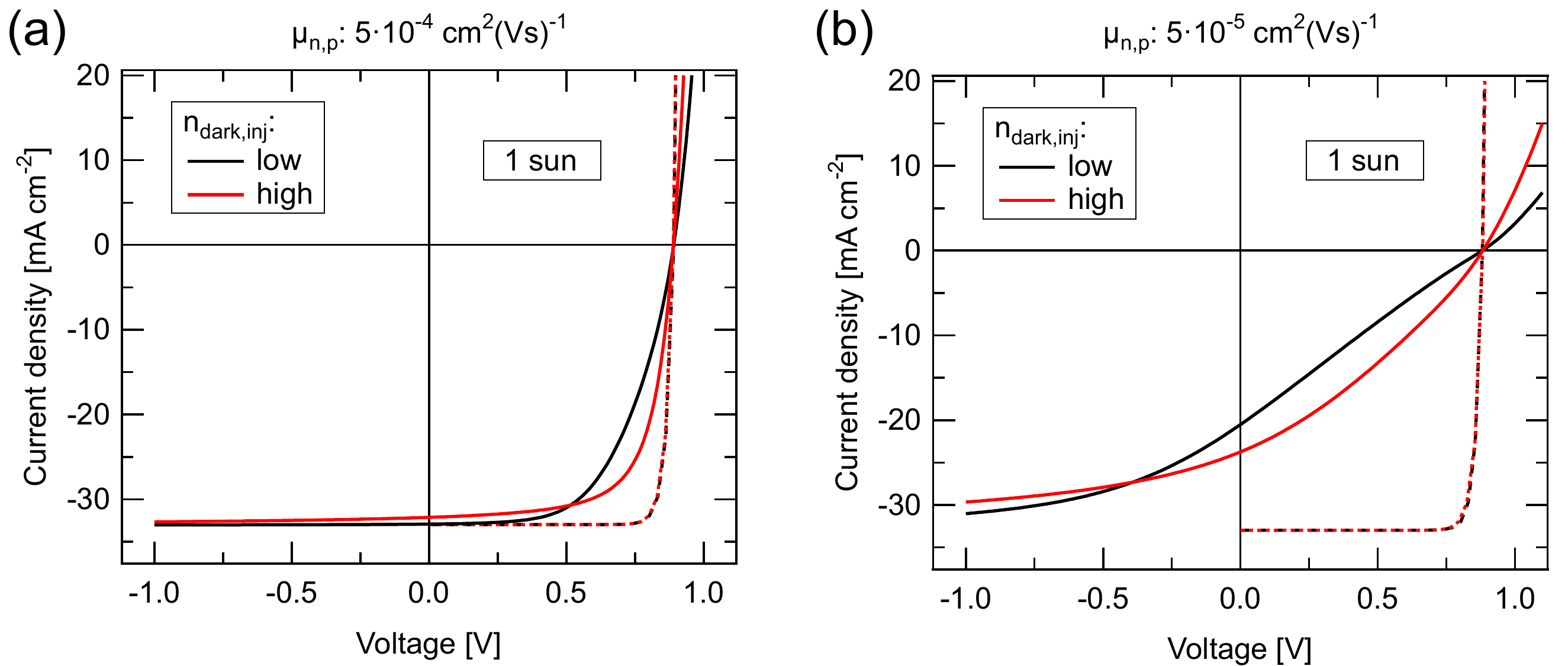}
    \caption{The impact of recombination with injected charge carriers $n_\mathrm{dark,inj}$, compared to transport resistance. Simulated $JV$ curves at 1~sun together with the suns-$\Voc$ curves (dashed lines), for charge carrier mobilities of (a) $5\cdot10^{-4}$ and (b)  $5\cdot10^{-5}\,\mathrm{cm^2(Vs)^{-1}}$, respectively. 
    }
    \label{fig:inj-JV}
\end{figure*}

Another mechanism that leads to fill factor losses -- beyond field-dependent photogeneration (section~\ref{sec:photogeneration}), nongeminate recombination and transport losses -- is the recombination of photogenerated charge carriers with injected (dark) charge carriers.\cite{dibb2011analysis,kniepert_conclusive_2014,deledalle2014understanding,wurfel_recombination_2019} This type of recombination is more pronounced for electrodes with small injection barriers, with a voltage dependence that differs from that of nongeminate recombination in the active layer bulk. Recombination with injected carriers is essentially a pseudo-first order process that occurs in the vicinity of the electrodes and requires two \enquote{ingredients}: (i) a large concentration of majority charge carriers which originates from Fermi level equilibration between the absorber and a metal or a highly doped transport material; (ii) an accumulation of minority charge carriers due to their low conductivity. Recombination with injected carriers can occur at the same time as transport resistance, and some of the conditions for both to occur are similar. However, they can in principle be distinguished, for instance by the light intensity dependence. 

\begin{figure*}[!tb]
    \centering
    \includegraphics[width=0.9\textwidth]{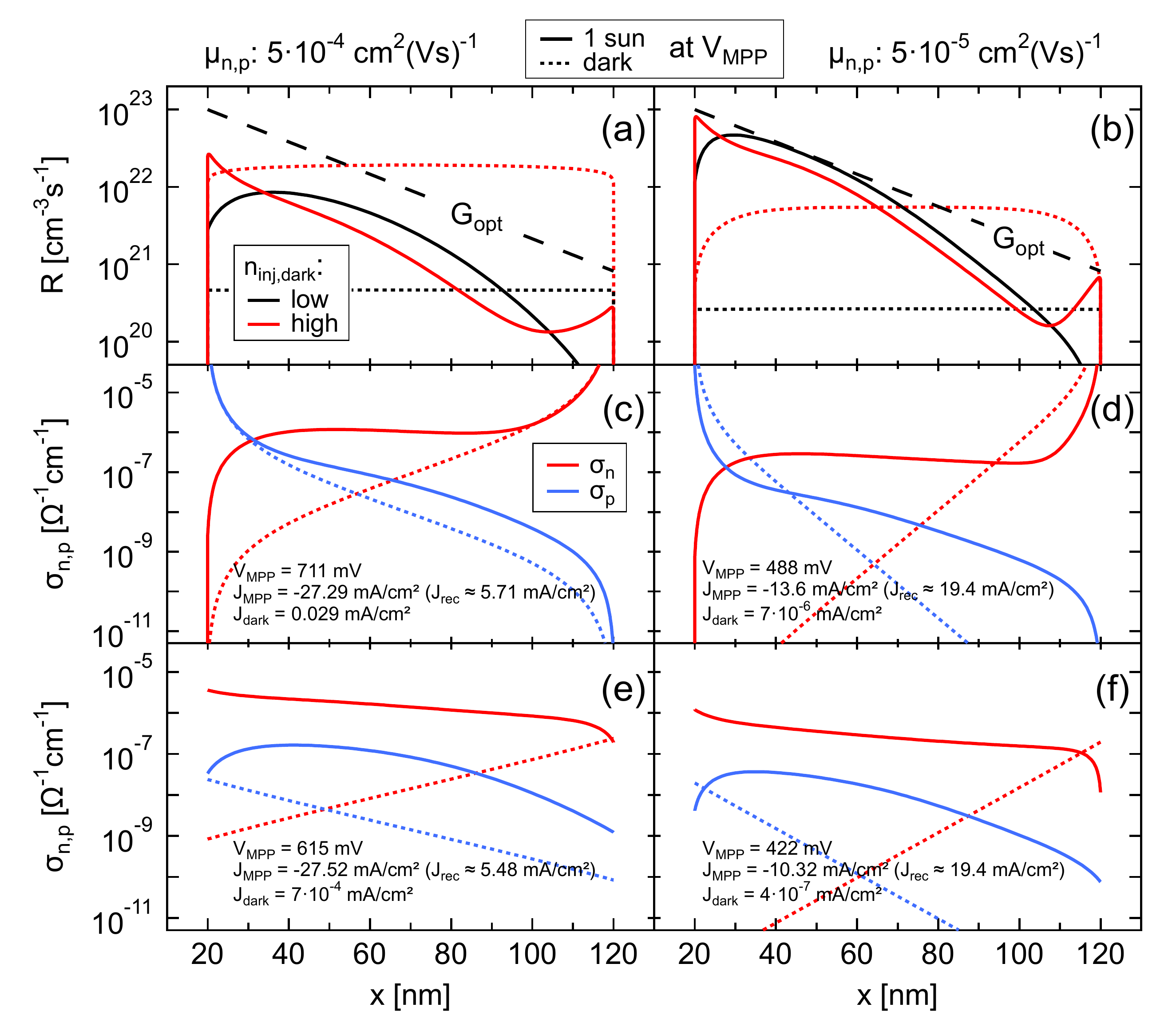}
    \caption{Spatial distribution of the recombination rate at $\Vmpp$ together with the optical generation rate for the two mobilities of (a,c,e) $5\cdot10^{-4}$ and (b, d, f) $5\cdot10^{-5}\,\mathrm{cm^2(Vs)^{-1}}$, respectively. The recombination rates in the dark (dashed lines) were scaled by $10^3$ in (a), and by $10^6$ in (b). (c)-(f) show the conductivities of electrons (red) and holes (blue) under illumination and in the dark with identical voltages applied.}
    \label{fig:inj-Rsigma}
\end{figure*}

In Figure~\ref{fig:inj-JV}(a) and (b) it can be seen that a high concentration of injected dark carriers $n_\mathrm{dark,inj}$ (red solid lines) -- achieved by low injection barriers -- improves the fill factor. This originates from the high concentration of majority charge carriers in the vicinity of their respective contact. While this high carrier concentration leads to an increased recombination rate in this region, the corresponding high majority carrier conductivity reduces the transport resistance. 
In order to get a feeling for the interplay between recombination with injected carriers and transport resistance losses, we chose perfectly selective charge extraction and distinguish two cases: The $JV$ curves in Figure~\ref{fig:inj-JV}(a) were simulated with balanced electron and hole mobilities of $5\cdot 10^{-4}\,\mathrm{ cm^2(Vs)^{-1}}$, whereas (b) shows simulations with factor 10 lower mobilities, with more pronounced transport losses. Note that the wide band gap electron and hole transport layers which extend from 0 to 20~nm and from 120 to 140~nm, respectively, are not shown for the sake of simplicity as generation and recombination do only occur in the photoactive layer.
In the case of the higher mobilities and a high value of $n_\mathrm{dark,inj}$, it can be seen that recombination at the maximum power point occurs predominantly near the contacts (red curve in Figure~\ref{fig:inj-Rsigma}(a). It is more pronounced near the hole contact (left) due to an inhomogeneous generation of charge carriers. In contrast, for a low value of $n_\mathrm{dark,inj}$ the recombination is bulk dominated. At short circuit, this effect causes the value of $\jsc$ in Figure~\ref{fig:inj-JV}(a) to be lower for a high value of $n_\mathrm{dark,inj}$, despite the higher fill factor. In Figure~\ref{fig:inj-JV}(b) it can be seen that the lower mobility does not only lead to lower fill factors, it also causes the value of $\jsc$ to be higher for the high value of $n_\mathrm{dark,inj}$. 
It can further be noted that the suns-$\Voc$ curves (dashed lines in (a) and (b)) are identical. This originates from the fact that at $\Voc$ recombination occurs rather homogeneously throughout the photoactive layer and thus the injected carrier concentration $n_\mathrm{dark,inj}$ plays a very minor role. 

In Figure~\ref{fig:inj-Rsigma}(a) and (b), the situation at the maximum power point is illustrated. Under illumination (solid lines), for a high value of $n_\mathrm{dark,inj}$ we observe more recombination at both ends of the photoactive layer, close to the electrodes, and less in the bulk. The dashed curves are the recombination rates at the same voltage (i.e., $\Vmpp$), but in the dark. The following three phenomena can be observed. (i) The recombination rates are \emph{much} smaller in the dark than under illumination (we multiplied the dark $R$ values in Figure~\ref{fig:inj-Rsigma}, by $10^3$ in (a) and $10^6$ in (b) to make them visible on this scale). This is in stark contrast to an ideal diode where the rates would be exactly the same, highlighting that the superposition principle is an ideal case. (ii) The dark recombination rates are spatially homogeneous and, (iii) for the high charge carrier mobilities in (a), $n_\mathrm{dark,inj}$ has a drastic impact on the recombination rate, whereas for the lower mobilities (b) it does not lead to any differences. Further, (c)-(f) show the corresponding conductivities of electrons (red) and holes (blue). Under illumination, they are significantly higher, especially for the electrons as they have to be transported over a larger distance. This increase in conductivity comes solely from the accumulation of charge carriers under illumination. Indicated are the simulated values of $\Vmpp$ and $\jmpp$ and also the recombination current densities under illumination (in brackets) and in the dark. The latter are, as already discussed above with respect to the recombination rates, orders of magnitude smaller due to the much smaller charge carrier concentrations.
This simulation study shows that the impact of dark injected charge carriers is two-fold: on the one hand, they increase the conductivities of the majority carriers exactly in those regions where their respective current is the largest (i.e., for electrons close to the electron contact and for holes close to the hole contact) and thus lead to a decrease of the effective transport resistance. On the other hand, they also induce an increased recombination rate in these regions, which manifests itself as a pseudo-first order loss. Therefore, injection of (dark) charge carriers can have positive and negative effects -- unlike the transport resistance, which is always a loss. Which of these two injection-induced effects -- increased conductivity or increased recombination -- is stronger depends on the transport properties and on the illumination intensity, that means, how many charge carriers have to be transported. 

A recent and insightful  approach\cite{sandberg2024diode} considered, for the first time, the impact of recombination with injected carriers on the $JV$ characteristics of OSC by an analytical model. The model also included the effect of the injected carriers on conductivity changes, as described in our simulation results above. The effect of the transport resistance of the photogenerated charge carriers, however, seems not to be accounted for. Generally, from our results, we think that both, transport resistance and recombination with injected carriers, have a distinct impact on the current--voltage characteristics. The recombination with injected carriers, as a pseudo-first order process, is essentially independent of light intensity and becomes less pronounced for thicker active layers as the relative fraction of the electrode region degreases. In contrast, transport losses increase with higher light intensity and thicker active layers. In the examples we presented above (Figure~\ref{fig:inj-JV}, the negative impact of transport losses dominated. The suns-$\Voc$ curve, however, contains no transport resistance losses and essentially no recombination with injected carriers. As a rule of thumb, if the FoM $\alpha$ changes with light intensity, with a dependence as described in section~\ref{sec:FoM}, then transport resistance losses dominate the overall fill factor losses. Similarly, the transport resistance losses dominate the difference between the illuminated $JV$ curve and the suns-$\Voc$ curve. Generally, further investigations to distinguish the effects of transport resistance and recombination with injected charge carriers are required.

\section{How to address the fill factor loss by the transport resistance?}\label{sec:outlook}

The transport resistance is caused by the interplay between poor charge transport and significant rate constants for recombination. In the following, we briefly discuss the various options to minimise these losses and thereby optimise the charge collection efficiency. We start by assembling the different material and device properties that influence the transport resistance. The transport resistance corresponds to a voltage drop over the active layer originating from the limited charge carrier conductivities in the latter. From the figure of merit $\alpha$, we see that a higher generation current increases the transport resistance, whereas a higher conductivity decreases it. The generation current is connected to the recombination current, as the mode of recombination determines the effective lifetime of the charge carriers and consequently their steady state density in the device. A longer effective lifetime (e.g., via a lower recombination prefactor $k$) has different effects depending on the bias point. At short circuit, it will reduce the amount of recombination at a given level of light intensity and for a given mobility. At open circuit, the recombination rate is fixed by the average generation rate and the longer lifetime will lead to a higher density of charge carriers at a given fixed generation (and therefore recombination) rate, thereby leading to a higher open-circuit voltage. In both situations, a longer lifetime will improve the solar cell parameters.

The most obvious method to minimise transport losses is to increase the effective conductivity. The lower of the electron and hole conductivities determines the transport resistance and limits the fill factor, as predicted by Ref.~\parencite{schiefer_determination_2014} Therefore, an optimised system should have balanced conductivities (section~\ref{sec:FoM}), which is in many cases well-approximated by balanced mobilities. 

Enhancing charge carrier mobility in the donor and acceptor is generally desirable for reducing transport losses. To this end, chemists focus on designing polymers and small molecules with rigid, coplanar backbones to improve molecular stacking, as well as side-chain engineering to tune interactions between molecules and between molecules and solvents, thereby promoting crystallinity.\cite{zhang_equally_2024} These strategies are effective for improving charge carrier mobility. One notable class of polymers exhibits temperature-dependent aggregation behaviour: they dissolve readily at elevated temperatures but form robust aggregates upon cooling to room temperature.\cite{hu_design_2017} This property enables the polymer phase to maintain high crystallinity in the solid film while avoiding excessively large crystalline domains that inhibit photogeneration (section~\ref{sec:meta-review}), overall improving charge carrier mobility and, ultimately, the fill factor. 

On the longer term, approaches for the high-mobility organic field-effect-transistor materials\cite{fratini2020charge} should be considered. Important open questions are whether the results are transferable\cite{xie_high-mobility_2024} to OPV despite the different working conditions -- lower carrier concentrations, high absorption coefficients, and the need to blend two materials? With the advances in high-throughput experiments\cite{torimtubun2024high} and the ability to computationally predict material properties from chemical structures,\cite{giannini2022charge} in principle new OPV materials with more ordered packing that lead to higher mobilities might be identified in the future.

A challenge is that these better packing or more crystalline organic semiconductors need to be compatible with the bulk heterojunction architecture -- at least from today's perspective. However, it might be that the requirement to blend two different materials, with different packing properties, together destroys the order of at least one of the constituents and, therefore, stands in the way of achieving a high order. This challenge can potentially be addressed by different approaches, among them the use of single-component donor--acceptor materials with a high-degree ordering,\cite{he_industrial_2022,li_molecular_2024} or even bilayers made from orthogonal solvents or by thermal evaporation, with the new high-order materials that also should have exciton diffusion lengths exceeding the donor and the acceptor layer thickness, respectively.

In addition to higher and balanced charge carrier mobilities, longer charge carrier lifetimes will be beneficial for the fill factor. The generation current as such is given by the illumination intensity, but the corresponding recombination rate can -- when the effective charge carrier lifetime is long -- lead to higher steady-state carrier concentrations. The minimisation of nonradiative recombination is generally an important tool to achieve PCE closer to the radiative efficiency limit. The nongeminate recombination in organic solar cells usually works by electrons and holes meeting to form charge transfer complexes that can recombine to the ground state. However, the higher the dissociation rate of charge transfer complexes back to charge carriers, the lower the net recombination rate. For simplicity of the argument, we assume that nongeminate recombination is given by a type of Langevin recombination. Then, a high photogeneration yield would decrease the Langevin prefactor,\cite{koster_origin_2005} leading to a lower recombination rate. Higher mobilities that were mentioned above to increase the conductivity would, for Langevin recombination, increase the recombination rate. This might be counter-productive, but then, the recombination is actually trap-assisted, and a higher material order that leads to higher mobilities will also lead to less traps: in the end, the recombination rate will decrease, not increase.

Another way to achieve higher conductivities is doping. However, early studies\cite{seemann_reversible_2011,schafferhans_oxygen_2010} have shown experimentally and by one-dimensional effective medium simulations that doping of one type of charge carrier in bulk heterojunction solar cells can negatively impact performance. The problem is that in a finely mixed donor--acceptor blend, no band bending occurs between the donor and acceptor domains. Consequently, doping to enhance the conductivity of one material impacts the other negatively, leading to increased recombination and overall performance loss. In principle, selective doping could potentially yield positive results by enabling band bending between the material phases in a system, where the donor and acceptor domain sizes are a few nanometres or larger. This was explored in an early study on transport resistance in organic bulk heterojunction solar cells, published in 2012.\cite{stelzl_modeling_2012} Two-dimensional device simulations were performed to model a solar cell with a bulk heterojunction active layer containing separated donor and acceptor phases, a scenario more realistic than the typically used effective medium approach. The authors suggested improving solar cell performance by individual p-doping of the donor and n-doping of the acceptor to minimise transport resistance. According to the two-dimensional simulations, both approaches positively influenced the fill factor and the overall performance. Selective doping has also made progress in improving carrier mobility and collection efficiency in non-fullerene-based OPVs by increasing the carrier concentration and charge mobility.\cite{xiong_revealing_2019}, or reshaping the internal potential distribution of the active layer, enhancing the extraction of low-mobility charge carriers.\cite{cui_eliminating_2023} However, implementing selective doping in practice is challenging. In addition to the potential issue of dopants from one material migrating into the other phase, the increased carrier concentrations resulting from selective doping may also promote exciton--charge carrier quenching.

Besides challenging long-term strategies, more straightforward device optimisation can be done already now. For the bulk heterojunction architecture, optimising the donor--acceptor nanomorphology is still the most accessible way to maximise charge collection. The well-intertwined bi-continuous fibrillar networks of donor--acceptor blends is considered one of the most suitable morphologies, which provides a good trade-off between sufficient interface for photogeneration and phase connectivity for charge carrier transport. From our meta-review (section~\ref{sec:meta-review}), we found that for some ternary blends, the degree of order is optimised, yielding a hierarchical morphology combined with elongated non-fullerene fibrils and finely sized polymer crystallites.\cite{zhu_single-junction_2022} Layer-by-layer deposition for preparing pseudo-bilayers has the advantage of maintaining the purity of the donor and acceptor domains as much as possible, forming a favourable vertical phase separation to promote charge collection by the corresponding electrode.\cite{jiang_pseudo-bilayer_2021} Both approaches have the potential to lower the transport resistance. Thinner active layers, potentially a challenge in roll-to-roll processing, might also be tested as a viable trade-off between charge photogeneration and charge collection. Generally, we recommend to perform further optimisation of the systems that were less in the focus of the OPV research community to minimise transport losses, such as inverted device structures, made from non-halogenated solvents or by thermal evaporation, without additives. This focussed effort will support the reduction of charge collection losses, allowing a successful and sustainable commercialisation of mass-produced OPV.

\section*{Acknowledgments}

We thank the Deutsche Forschungsgemeinschaft (DFG) for funding this work (Research Unit FOR~5387 POPULAR, project no.~461909888). CD is grateful to Oskar Sandberg for the interesting discussions. TK acknowledges funding by the Helmholtz Association via the POF IV funding as well as via the SolarTap project. TK further thanks the Ministry of Economic Affairs, Industry, Climate Action and Energy of the State of North Rhine-Westphalia for funding via the ENFA (\enquote{Entwicklung effizienter ternärer NFA-basierter organischer Photovoltaik durch Machine-Learning Methoden}) project. UW acknowledges funding from the German Federal Ministry for Economic Affairs and Energy (FKz.~03EE1170 — StabilO-PV).

\bibliographystyle{apsrev4-2}
\bibliography{references}

\end{document}